# High-throughput ensemble characterization of individual core-shell nanoparticles with quantitative 3D density maps from XFEL single-particle imaging

Do Hyung Cho[1,†], Zhou Shen[2,†], Yungok Ihm[3], Dae Han Wi[4], Chulho Jung[1], Daewoong Nam[5], Sangsoo Kim[5], Sang-Youn Park[5], Kyung Sook Kim[5], Daeho Sung[1], Heemin Lee[1], Jae-Yong Shin[1], Junha Hwang[1], Sung-Yun Lee[1], Su Yong Lee[5], Sang Woo Han[4], Do Young Noh[6,7], N. Duane Loh[2,8*], Changyong Song[1,9*]

*[1]Department of Physics & Photon Science Center, POSTECH, Pohang 37673, Korea, [2]Department of Physics, National University of Singapore, Singapore 117551, [3]Department of Chemistry, POSTECH, Pohang 37673, Korea, [4]Center for Nanotectonics, Department of Chemistry and KI for the NanoCentury, KAIST, Daejeon 34141, Korea, [5]Pohang Accelerator Laboratory, POSTECH, Pohang 37673, Korea, [6]Department of Physics and Photon Science, Gwangju Institute of Science and Technology, Gwangju 61005, Korea, [7]Institute for Basic Science (IBS), Daejeon 34126, Korea, [8]Department of Biological Sciences, National University of Singapore, Singapore 117557, [9]Asia Pacific Center for Theoretical Physics (APCTP), POSTECH, Pohang 37673, Korea*

† *These authors contributed equally.*

**Corresponding authors: duaneloh@nus.edu.sg & cysong@postech.ac.kr*

**The structures, as building-blocks for designing functional nanomaterials, have fueled the development of versatile nanoprobes to understand local structures of noncrystalline specimens. Progresses in analyzing structures of individual specimens with atomic scale accuracy have been notable recently. In**

**most cases, however, only a limited number of specimens are inspected lacking statistics to represent the systems with structural inhomogeneity. Here, by employing single-particle imaging with X-ray free electron lasers and new algorithm for multiple-model 3D imaging, we succeeded in investigating several thousand specimens in a couple of hours, and identified intrinsic heterogeneities with 3D structures. Quantitative analysis has unveiled 3D morphology, facet indices and elastic strains. The 3D elastic energy distribution is further corroborated by molecular dynamics simulations to gain mechanical insight at atomic level. This work establishes a new route to high-throughput characterization of individual specimens in large ensembles, hence overcoming statistical deficiency while providing quantitative information at the nanoscale.**

Materials' functions are contingent on their structural properties[1,2]. This contingency has spurred the research on functional nanomaterials by designing structures that accommodate desired functions[3]. In this capacity, high-resolution structural probes are critical for detecting emergent functional properties induced by local structural motifs[4-6]. Detecting these motifs requires local structural characterization to image large numbers of specimens or large regions. Examples include reaction kinetics from *in situ* characterization and catalytic function of nanocrystals from 3D mapping of elastic strains, where such local probes using electrons or X-rays allow us to discover and study structure-function relations[6-14]. However, most of these local 3D probes only operate in a low-sampling mode requiring days of measurement time for each specimen. As such, inspecting only a few specimens makes it difficult to understand the whole system with statistical confidence. Such limited measurements can be a critical shortcoming, as the intrinsic structural inhomogeneity is noticeable even for nanomaterials prepared under identical growth conditions[15].

Here we demonstrate a high-throughput nanoscale characterization of an ensemble of individual specimens using X-ray free electron laser (XFEL) single-particle imaging. Unlike X-ray based 3D characterization methods for samples too thick for electron microscopy, we interrogated thousands of specimens in a couple of hours. The whole 3D elastic strain energy distribution and strain fields unveiled from the nanoscale 3D density maps provide structural insight into the functional activity of the materials; elastic strains play key roles in the catalytic reactivity, for instance, with the strain-induced modification of the energy bands[16]. Furthermore, the XFEL provides the advantage for studying native structures with imperceptible radiation damage, where diffraction signals from femtosecond X-ray pulses outrun radiation induced structure changes[17]. Thereafter, statistically-significant structural motifs were recovered with advancements in applying machine-learning for unsupervised data screening and 3D imaging. Such high-throughput statistical learning of structural classes holds strong potential for studying ensembles of heterogeneous and dynamic structures[11,18,19].

We have investigated core-shell nanoparticles (NPs), Au@$TiO_2$, as a model NP system (Methods). Each nanoparticle comprises a concave high-index-faceted trisoctahedral (TOH) Au nanocrystal (NC) of 120 nm diameter, encapsulated by a spherical shell of 70 nm thick $TiO_2$ layer. This core-shell NP system has attracted interest as light harvesting agents with surface plasmonic resonance enhancements, and the knowledge on its morphological and mechanical feature is essential for improving conversion efficiency[2,20]. Additionally, the nanocrystals with high-index facets have attracted continued interest as good functional materials for catalyst, partly due to their facet-dependent catalytic activity including molecular adsorption[21].

Single-particle 3D imaging experiments were performed at PAL-XFEL (Pohang Accelerator Laboratory-XFEL) with the specimens loaded on thin $Si_3N_4$ membranes employing the fixed-target scheme (Methods)[22,23]. We collected single-pulse two-dimensional (2D) diffraction patterns of specimens at random orientations by translating the membrane window across the focused XFEL beam (Fig. 1a and Methods). Statistically meaningful characterization that uncovers structural features present in a whole ensemble of specimens necessitates collecting a large number of diffraction patterns enough to capture the structural diversity, which became realized through single-pulse data collection using the XFEL. We collected approximately 120,000 diffraction patterns in ~ 100 minutes with the XFEL running at 30 Hz.

An equally essential requirement for the ensemble characterization is the capability of inspecting such big data and sorting out meaningful ones for in-depth analysis. We have employed a machine-learning algorithm for automatic data screening to filter out patterns from multiple-particle hits, weak signals without distinct speckles, or significantly different structures[24,25]. The automatic sorting of the diffraction patterns was made by projecting the data onto the two principal components (PCs) of a feature space reduced using the *Xception* deep neural network (Fig. 1b and Methods)[26]. From the 120,000 (120K) single-pulse diffraction patterns, finally 1,176 patterns were saved for 3D reconstructions. The saved data were clearly distinguished in the PC analysis with the 1,176 single-particle patterns clustered at the center of $PC_1$, whilst the 120K points were distributed widely (Fig. 1b). The intrinsic structural heterogeneity of the Au@$TiO_2$ NPs is reflected on these 1,176 diffraction patterns, which were classified and assembled separately for 3D images to represent types of inhomogeneous structures present in the specimens.

Use of the fixed-target sample loading for the XFEL single-particle imaging, first demonstrated here, has helped to significantly improve the data collection efficiency. Experimentally achieved rates of single-particle hit events and multiple-particle hit were displayed (open rectangles), which showed good agreement (> 98 %) with the theoretical estimation for single (blue solid line) and multiple-particle (red) hits at the areal particle number density of ~ 0.09 $\mu m^{-2}$ (Fig. 1c). The solid line drawn in black color shows the total hit rate, and the 95 % total hit rates noted during the experiment is consistent with the theoretical expectation[24]. About one out of fourteen single-particle patterns displayed strong intensities, which were used for 3D reconstructions. These high-quality single-particle patterns, which corresponded to ~ 1 % of total hits, were obtained out of $1.6\times10^5$ NPs using this fixed-target scheme, which is ~ 2 orders higher data collection rate than a state-of-the-art aerosol injector provides[27].

The coherent X-ray diffraction pattern, collected at a Fraunhofer diffraction regime, is equivalent to the square-modulus of Fourier transform of optically thin specimen's electron density projected along the X-ray beam direction. With the random orientations of the specimens relative to X-ray pulses, each single-particle diffraction pattern corresponds to a random 2D central Ewald-sphere section of a core-shell nanoparticle's 3D diffraction volume (Fig. 1a). Accordingly, single-particle diffraction patterns from different structures of the Au@$TiO_2$ NPs belong to different 3D diffraction volumes representing the inherent structural polydispersity (Fig. 1d).

The unprecedented advantage of XFEL single-pulse diffraction in high-resolution investigation of functional nanomaterials was explicitly verified from the direct comparison of the diffraction data obtained from similar experiments using synchrotron X-rays (Fig. 2 and Methods). XFEL single-pulse diffraction patterns displayed spatial resolutions superior to those from synchrotrons for similarly-oriented individual NPs (Fig. 2a and b). In experiments using synchrotron X-rays, high spatial resolution is typically achieved by increasing X-ray exposure time[28]. We, however, showed that resolution became intrinsically limited by the X-ray radiation damage that was accumulated during much longer exposure times at synchrotrons (Fig. 2c-e). This unequivocally demonstrates that, for nanometer-resolution imaging, femtosecond X-ray pulses are critical in overcoming resolution limits due to the radiation damage[12,17].

The 3D diffraction volume was assembled by identifying orientations of 2D patterns employing the Expand-Maximize-Compress (EMC) algorithm[29,30]. The EMC algorithm reconstructs the 3D diffraction pattern using an expectation maximization scheme to infer the latent variables such as orientations and local incident photon fluence without imposing *a priori* information about specimen's symmetry[31]. We identified types of structural heterogeneity using a statistical approach. To identify the desired subset of all 2D patterns, we reconstructed multiple 3D volumes ('models') simultaneously from the full set of 2D patterns *de novo* [32]. This multiple-model EMC (mmEMC) essentially partitions all 2D patterns into overlapping subsets that maximize the posterior probability of the reconstructed 3D models (Methods and Supplementary Methods). Repeated mmEMC reconstructions from random initialization indicated that the 2D patterns are most consistent with four statistically significant 3D models indicating that intrinsic structural heterogeneity in Au@$TiO_2$ NPs can be classified into four discernible types (Fig. 1d and Supplementary Video 1).

The 3D diffraction volumes recovered by the mmEMC were examined in at least three independent ways: comparison both with simulated structure and with experimentally measured projection images and symmetry analysis. The 3D diffraction volume of the type-I NP is displayed using color-coded contour map (Fig. 3a). We extracted 2D sections from this diffraction volume along the three major crystallographic directions of [100], [110] and [111] (upper panel in Fig. 3b), and compared them with the simulated patterns generated using a nominal core-shell structure (lower panel in Fig. 3b).

By performing numerical phase retrieval of the assembled 3D diffraction volume, we obtained the 3D image of the specimen (Fig. 3c, Methods, Supplementary Methods and Supplementary Video 2)[33]. Projection images of the 3D structures were also compared with the 2D images from direct reconstruction of the single-pulse 2D diffraction patterns as well as electron microscope images validating the structure (Extended Data Fig.1). Inhomogeneous density distribution in the NP was observed clearly in numerically sectioned images. The 3D image resolution was estimated to be ~ 20 nm (Supplementary Fig. 5), which includes resolution loss from the structural heterogeneity between different NPs whose diffraction patterns were used to reconstruct the type-I volume. Despite this heterogeneity, the characteristic concave TOH structure was identified in the reconstructed 3D structure from the views along the [100], [110] and [111] directions (upper panel in Fig. 3d). The images were compared with the simulated images of ideal TOH core and a spherical shell (lower panel in Fig. 3d). The accuracy of this 3D structure was further verified by the strong signatures of octahedral symmetry in the diffraction volumes recovered by mmEMC (Fig. 3e and Methods).

There are visual differences among the types I-III NPs that we can characterize quantitatively; we excluded the type-IV NP that displayed significantly distorted morphologies. Assuming a spherical $TiO_2$ shell without distinct structures, the following analysis was focused on the TOH Au NC. First, we calculated Gaussian curvatures to quantify their 3D morphologies (Fig. 4a, d, g, Methods and Supplementary Video 3). The regions colored in red had 40% smaller radius of curvature than a sphere of a similar size, indicating local extrusion from the surface having an average radius of curvature of 60 nm. The curvatures at the 8 pyramidal vertices (red open circles in the trisoctahedron shown in Fig.1a), extruded from the center of octahedral planes, were typically larger than the 6 octahedral vertices by 1.3~1.8 times. The type-II NC displayed smooth edges with smaller curvature compared to the type-I and type-III NCs (Fig. 4a, d and g).

The 3D distribution of local curvatures was displayed using angular plots (Fig. 4b, e and h). Morphological distortions of the NCs was evident when comparing the connections between local curvature maxima (red lines) with those from an ideal TOH structure (broken white lines). Once the TOH vertices of type I-III volumes were registered to those on the ideal TOH, their facets can be indexed (Fig. 3c, f, i, Methods, Supplementary Methods and Supplementary Video 4). The NC surfaces were mainly composed of {331} planes along with the {221} and {552} for the type-I structure (Fig. 4c), consistent with facet indices reported frequently for the TOH nanoparticles[34,35]. Similar facet surfaces were noted for other types of structures with the common appearance of the {331} planes, but facet surfaces with higher indices were also present.

Next, we have analyzed inhomogeneous mass density observed in the Au NC. With the crystalline nature of the Au core, the density variation results from a local contraction and expansion of the lattice due to the accumulation of local strain. We quantified the strain directly from the 3D density map to estimate elastic strain energy using the bulk modulus of the Au (Fig. 5a, d, g, Methods, Supplementary Methods and Supplementary Video 5). Compared to the rest of the crystal, this energy in type-I Au NC was higher at the TOH vertices as well as along edges of the octahedron where the two pyramids meet suggesting the mechanical instability of the vertices and the edges (Fig. 5a). Consequently, the maximal energy gain by the strain reached to 0.5 eV per atom that is ~ 13 % of Au cohesive energy. This corresponds to the lattice distortion as large as 0.28 Å, or ~ 7 % of the nominal interatomic distance.

Further we characterized the 3D distribution of the strain field within the Au NC, and displayed it using arrows guiding the direction of the steepest increase in the elastic strain energy (Fig. 5b, e and h). This is superimposed with a 3D contour map showing comparable strain energy surfaces with the same degree of volume dilation at ± 10 % level of the average value. The dilation map with positive values (blue) indicated volume expansion and negative (red) for contraction. A cross-sectional view displayed the strain field more clearly with the arrows indicating the strain present near the sectioned plane (Fig. 5c, f, i and Supplementary Video 6). The compressive strain is concentrated at the TOH vertices of the type-I Au NC, to be released by moving toward the edge of the octahedron connecting two pyramids. It shared the features with the knowledge that bond lengths are strongly contracted for the atoms at edges and corners of the surface with different coordination numbers[14]. The reconstructed image of type-I NC helped to further refine this to accompany lattice relaxation at the edges and compression at the corners. This strain field with the loci of compressed and expanded regions was similarly observed but with rather asymmetric distribution for type-II and type-III Au NCs.

We then compared this strain energy map by calculating the elastic energy both for the type-I NC and an ideal TOH Au NC model structure using all-atom molecular dynamics (MD) simulations (Fig. 5j, k, Methods and Supplementary Methods). The former was constructed by filling the volume of type-I NC with face-centered-cubic Au lattice of matching density, which was then relaxed by MD simulations to release artificial strains formed during this construction (Fig. 5j and Methods). Similarly, the ideal TOH crystal was also relaxed (Fig. 5k and Supplementary Video 7). The elastic energy calculated for the model structure showed good agreement with the experimental result validating the strain energy estimated directly from the 3D density (Fig. 5j). For the ideal trisoctahedron, higher elastic energy was noted mostly at the pyramidal vertices. However, the relaxation of the lattice, observed at the edges of the octahedron in the experimental results, was not as significantly accompanied in the ideal trisoctahedron (Fig. 5k). These differences between reconstructed and ideal structures highlight the value of recovering actual structural motifs present in our ensemble over presuming ideality.

In summary, we have identified persistent nanoscale 3D structural motifs within an ensemble comprising thousands of individual Au@$TiO_2$ core-shell NPs. This high-throughput and individual nanoscale characterization was realized using XFEL single-particle 3D imaging with newly developed imaging algorithms and analysis methods. The femtosecond XFEL pulses, as explicitly demonstrated by directly comparing the diffraction patterns from synchrotrons, provided distinct advantage in revealing the structure at a resolution beyond the radiation-damage limit. With these, in-depth analysis of 3D morphology, mass density, strain energy distribution and strain fields was successfully undertaken with minimal disturbance to the specimens. The 3D morphology and strain distributions learned from the present study promoted the understanding on materials functions with the direct impact of strains on the catalytic reactivity; adsorption and desorption rate of the oxygen or CO molecules are contingent on the metal *d*-band states that is affected by the interatomic spacing at the metal surface[16].

This high-throughput ensemble characterization workflow is immediately applicable for a wide range of functional materials with the new leverage on *in situ* investigation of ultrafast dynamics facilitated by femtosecond XFEL pulses. We expect to observe paradigm shifting changes in nanostructure analysis with this XFEL ensemble characterization method, which will be further enhanced by upcoming megahertz (MHz) XFELs that increase the sample probe rates by a thousand-fold[36-39]. Such multiple orders higher sampling rates from the MHz-XFEL will bring an immediate impact to the ultrafast dynamics research with the feasibility to explore additional parameter space such as the time, which will realize the four-dimensional high-throughput ensemble investigation of individual specimens at femtosecond temporal and nanoscale spatial resolutions[11,40].

**References**

1 Billinge, S. J. L. & Levin, I. The Problem with Determining Atomic Structure at the Nanoscale. *Science* **316**, 561-565 (2007).

2 Luo, M. & Guo, S. Strain-controlled electrocatalysis on multimetallic nanomaterials. *Nat. Rev. Mater.* **2**, 17059 (2017).

3 Oh, M. H. *et al.* Design and synthesis of multigrain nanocrystals via geometric misfit strain. *Nature* **577**, 359-363 (2020).

4 Miao, J., Ishikawa, T., Robinson, I. K. & Murnane, M. M. Beyond crystallography: Diffractive imaging using coherent x-ray light sources. *Science* **348**, 530 (2015).

5 Pfeifer, M. A., Williams, G. J., Vartanyants, I. A., Harder, R. & Robinson, I. K. Three-dimensional mapping of a deformation field inside a nanocrystal. *Nature* **442**, 63-66 (2006).

6 Robinson, I. & Harder, R. Coherent X-ray diffraction imaging of strain at the nanoscale. *Nat. Mater.* **8**, 291-298 (2009).

7 Yang, Y. *et al.* Deciphering chemical order/disorder and material properties at the single-atom level. *Nature* **542**, 75-79 (2017).

8 Clark, J. N. *et al.* Ultrafast Three-Dimensional Imaging of Lattice Dynamics in Individual Gold Nanocrystals. *Science* **341**, 56-59 (2013).

9 Cha, W. *et al.* Core–shell strain structure of zeolite microcrystals. *Nat. Mater.* **12**, 729-734 (2013).

10 Kim, D. *et al.* Defect Dynamics at a Single Pt Nanoparticle during Catalytic Oxidation. *Nano Letters* **19**, 5044-5052 (2019).

11 Ihm, Y. *et al.* Direct observation of picosecond melting and disintegration of metallic nanoparticles. *Nat. Commun.* **10**, 2411 (2019).

12 Gallagher-Jones, M. *et al.* Macromolecular structures probed by combining single-shot free-electron laser diffraction with synchrotron coherent X-ray imaging. *Nat. Commun.* **5**, 3798 (2014).

13 Chen, C.-C. *et al.* Three-dimensional imaging of dislocations in a nanoparticle at atomic resolution. *Nature* **496**, 74-77 (2013).

14 Huang, W. J. *et al.* Coordination-dependent surface atomic contraction in nanocrystals revealed by coherent diffraction. *Nat. Mater.* **7**, 308-313 (2008).

15 Kim, B. H. *et al.* Critical differences in 3D atomic structure of individual ligand-protected nanocrystals in solution. *Science* **368**, 60 (2020).

16 Mavrikakis, M., Hammer, B. & Nørskov, J. K. Effect of Strain on the Reactivity of Metal Surfaces. *Phys. Rev. Lett.* **81**, 2819-2822 (1998).

17 Neutze, R., Wouts, R., van der Spoel, D., Weckert, E. & Hajdu, J. Potential for biomolecular imaging with femtosecond X-ray pulses. *Nature* **406**, 752-757 (2000).

18 Nango, E. *et al.* A three-dimensional movie of structural changes in bacteriorhodopsin. *Science* **354**, 1552 (2016).

19 Loh, N. D. *et al.* Fractal morphology, imaging and mass spectrometry of single aerosol particles in flight. *Nature* **486**, 513-517 (2012).

20 Liu, W.-L. *et al.* The influence of shell thickness of Au@TiO2 core–shell nanoparticles on the plasmonic enhancement effect in dye-sensitized solar cells. *Nanoscale* **5**, 7953-7962 (2013).

21 Huang, M. H., Rej, S. & Hsu, S.-C. Facet-dependent properties of polyhedral nanocrystals. *Chem. Commun.* **50**, 1634-1644 (2014).

22 Kang, H.-S. *et al.* Hard X-ray free-electron laser with femtosecond-scale timing jitter. *Nat. Photon.* **11**, 708-713 (2017).

23 Nam, D. *et al.* Fixed target single-shot imaging of nanostructures using thin solid membranes at SACLA. *J. Phys. B-At. Mol. Opt.* **49**, 034008 (2016).

24 Park, J., Joti, Y., Ishikawa, T. & Song, C. Monte Carlo study for optimal conditions in single-shot imaging with femtosecond x-ray laser pulses. *App. Phys. Lett.* **103**, 264101 (2013).

25 Rose, M. *et al.* Single-particle imaging without symmetry constraints at an X-ray free-electron laser. *IUCrJ* **5**, 727-736 (2018).

26 Chollet, F. Xception: Deep learning with depthwise separable convolutions. in *The IEEE Conference on Computer Vision and Pattern Recognition.* 1251-1258 (2017).

27 Bielecki, J. *et al.* Electrospray sample injection for single-particle imaging with x-ray lasers. *Sci. Adv.* **5**, eaav8801 (2019).

28 Jung, C. *et al.* Structural Investigation of Single Specimens with a Femtosecond X-Ray Laser: Routes to Signal-to-Noise Ratio Enhancement. *Phys. Rev. Applied* **13**, 064045 (2020).

29 Loh, N.-T. D. & Elser, V. Reconstruction algorithm for single-particle diffraction imaging experiments. *Phys. Rev. E* **80**, 026705 (2009).

30 Ekeberg, T. *et al.* Three-Dimensional Reconstruction of the Giant Mimivirus Particle with an X-Ray Free-Electron Laser. *Phys. Rev. Lett* **114**, 098102 (2015).

31 Fung, R., Shneerson, V., Saldin, D. K. & Ourmazd, A. Structure from fleeting illumination of faint spinning objects in flight. *Nat. Phys.* **5**, 64-67 (2009).

32 Ayyer, K., Lan, T.-Y., Elser, V. & Loh, N. D. Dragonfly: an implementation of the expand-maximize-compress algorithm for single-particle imaging. *J. App. Cryst.* **49**, 1320-1335 (2016).

33 Loh, N.-T. D., Eisebitt, S., Flewett, S. & Elser, V. Recovering magnetization distributions from their noisy diffraction data. *Phys. Rev. E* **82**, 061128 (2010).

34 Ma, Y. *et al.* Synthesis of Trisoctahedral Gold Nanocrystals with Exposed High-Index Facets by a Facile Chemical Method. *Angew. Chem, Int. Ed.* **47**, 8901-8904 (2008).

35 Yu, Y., Zhang, Q., Lu, X. & Lee, J. Y. Seed-Mediated Synthesis of Monodisperse Concave Trisoctahedral Gold Nanocrystals with Controllable Sizes. *J. Phys. Chem. C* **114**, 11119-11126 (2010).

36 Sobolev, E. *et al.* Megahertz single-particle imaging at the European XFEL. *Commun. Phys.* **3**, 97 (2020).

37 Decking, W. *et al.* A MHz-repetition-rate hard X-ray free-electron laser driven by a superconducting linear accelerator. *Nat. Photon.* **14**, 391-397 (2020).

38 Galayda, J. The LCLS-II: a high power upgrade to the LCLS. in *Proceedings of the 9th International Particle Accelerator Conference.* 18-23 (JACoW Publishing, 2018).

39 Zhu, Z. Y., Zhao, Z. T., Wang, D., Yin, L. X. & Yang, Z. H. SCLF: an 8-GeV CW SCRF linac-based X-ray FEL facility in Shanghai. in *Proceedings of the 38th International Free Electron Laser Conference.* 182-184 (JACoW Publishing, 2017).

40 Zewail, A. H. Four-dimensional electron microscopy. *Science* **328**, 187-193 (2010).

41 Lee, H. *et al.* Characterizing the intrinsic properties of individual XFEL pulses via single-particle diffraction. *J. Syn. Rad.* **27**, 17-24 (2020).

42 Pham, M., Yin, P., Rana, A., Osher, S. & Miao, J. Generalized proximal smoothing (GPS) for phase retrieval. *Opt. Express* **27**, 2792-2808 (2019).

43 Plimpton, S. Fast Parallel Algorithms for Short-Range Molecular Dynamics. *J. Comp. Phys.* **117**, 1-19 (1995).

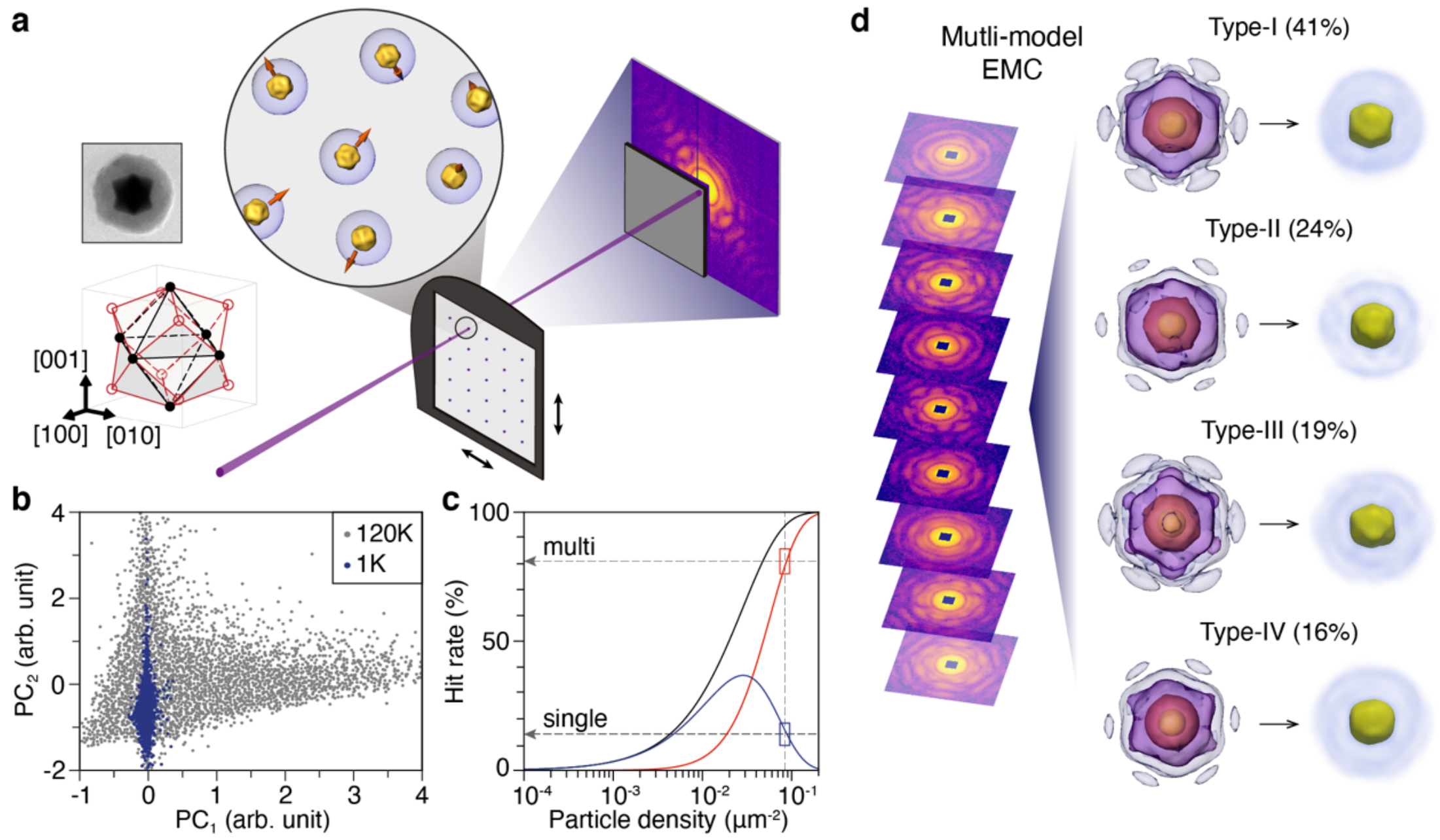

**Fig. 1| High-throughput ensemble characterization of individual nanoparticles with femtosecond X-ray single-particle 3D imaging. a,** Single-pulse diffraction patterns are obtained for NPs dispersed on thin $Si_3N_4$ membrane in random orientations. The arrows on the particles visualize the random orientation in the zoom-in view. Crystallographic orientation of the TOH Au NC is shown in the inset. **b,** Machine-learning based single-particle classification. The gray dots represent the whole 120,000 diffraction patterns (120K) and the blue dots for 1,176 (1K) diffraction patterns from single NPs. **c**, Experimentally achieved hit rates (open rectangles) showing good agreements with the theoretical estimation for single (blue solid line) and multiple-particle (red) hits. **d,** Single-pulse diffraction patterns collected from individual specimens with intrinsic inhomogeneity are classified into four structural types (I-IV) using the multi-model EMC (relative occupancy indicated as a percentage). Diffraction patterns belonging to the same structure type are assembled into their respective 3D diffraction volumes, on which phases are retrieved to obtain 3D real images.

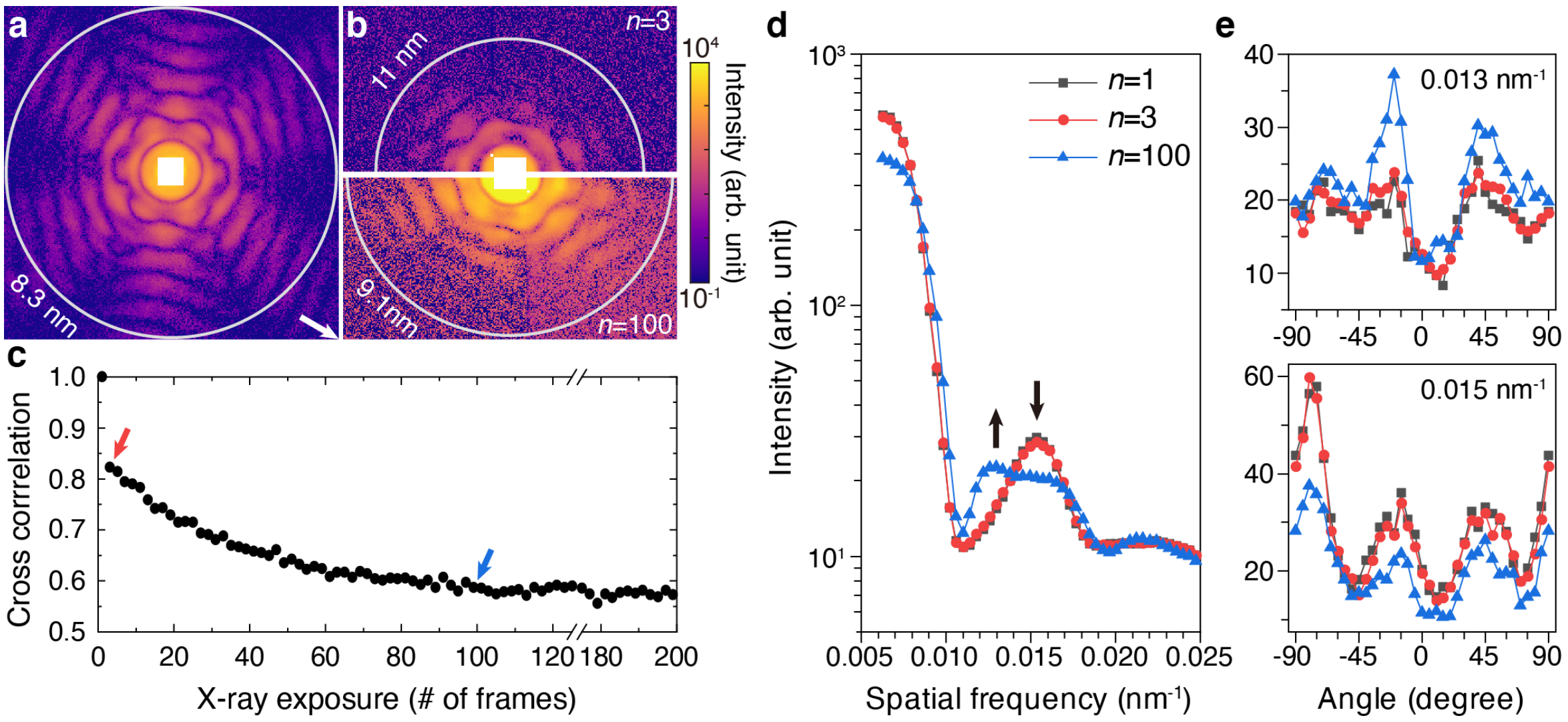

**Fig. 2| Single-particle coherent diffraction using synchrotron X-rays and radiation induced sample damage. a,** XFEL single-pulse diffraction with the signal reaching to the resolution of 8.3 nm (white circle) in the edge and 5.9 nm at the corner (white arrow). **b,** Diffraction using synchrotron X-rays at a comparable level of X-ray dose used in XFEL in (a) with the scattering signal faded out at ~ 11 nm (white circle, upper half panel). The diffraction pattern by accumulating 100 frames without meaningful gain in resolution hampered by radiation induced sample damage (lower half panel). **c,** The cross correlation (CC) of each diffraction pattern between the first exposure and *n*th exposures with low CC from the early stage. **d,** Radial intensity plot shows the alteration in the diffraction pattern after the accumulated dose of X-ray radiation. **e,** The diffraction intensities were compared for the data collected from the 1$^{st}$, 3$^{rd}$ and 100$^{th}$ exposure at selected spatial frequencies of 0.013 and 0.015 nm$^{-1}$ along the azimuthal angle displaying notable variation of the pattern with the repeated exposure to the synchrotron X-rays.

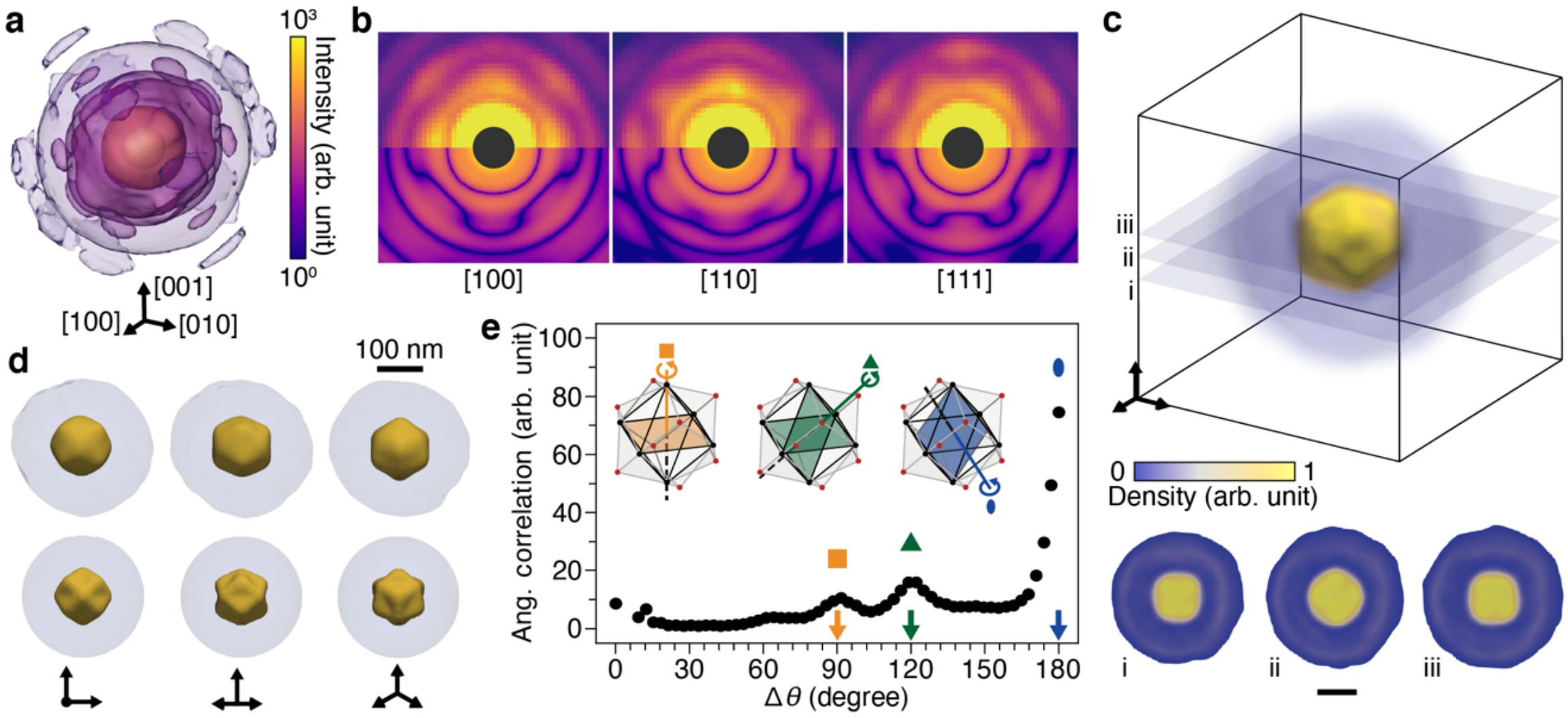

**Fig. 3| Nanoscale 3D structure of type-I Au@$TiO_2$ core-shell nanoparticle. a,** Assembled 3D diffraction volume shown using a contour map. **b,** 2D sections extracted from this assembled 3D volume in (a) along the [100], [110] and [111] directions (upper half panel) are compared with the simulated diffraction patterns (lower half panel). **c,** Reconstructed 3D electron density map after the phase retrieval of the assembled 3D diffraction volume in (a). Cross-sectioned images display internal density distributions and boundaries of the TOH Au NC. The scale bar is 100 nm. **d,** Perspective views of the 3D structure along the crystallographic directions of [100], [110] and [111] (upper panel) are compared with the nominal structure in good agreement (lower panel). **e,** Angular correlation of diffraction patterns' orientations with posterior probability distribution confirming TOH symmetry of the sample.

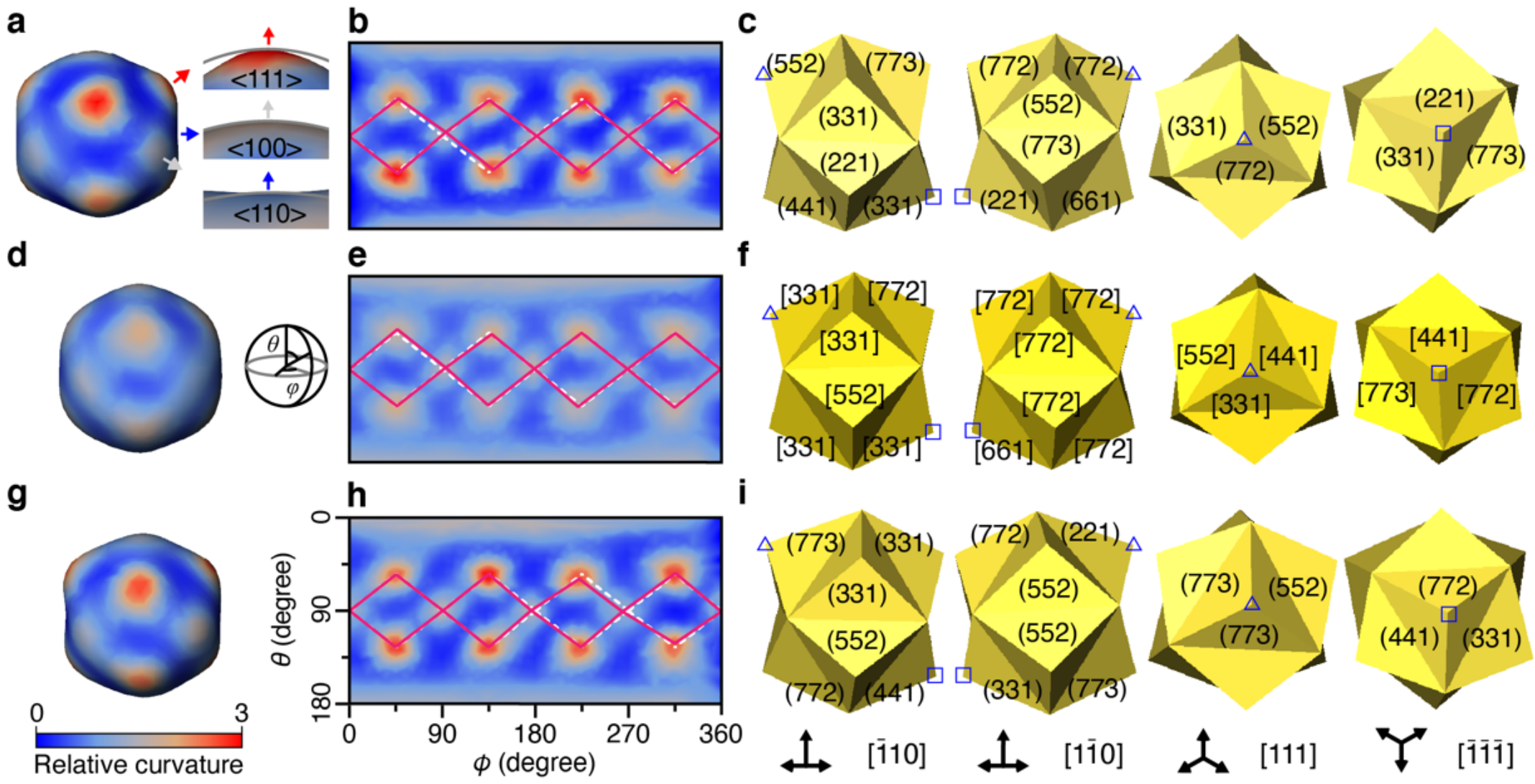

**Fig. 4| 3D morphology and identification of facet indices**. **a,** Surface morphology of the type-I NP was shown with local curvature in color scale. Insets show the magnified view of the surface morphology. Locally extruded regions with higher curvature (in red) are located at the vertices of trisoctahedron. **b,** Surface map of local curvature is displayed using angular plot confirming the TOH symmetry. The solid red lines connect the local curvature maxima, which appear slightly displaced from the position expected for the ideal trisoctahedron (broken white lines). **c,** Facet indices were identified from the 3D morphology. The 3D morphology is best matched with high-index-faceted trisoctahedron that is consists of {331}, {221}, {552}, and {772} planes mostly. Facet planes are shown for different orientation, with the viewing direction indicated, for full 3D views. Similar plots of 3D morphology (**d**), angular plot (**e**), and facet indices (**f**) for type-II NP and type-III NP (**g, h** and **i**).

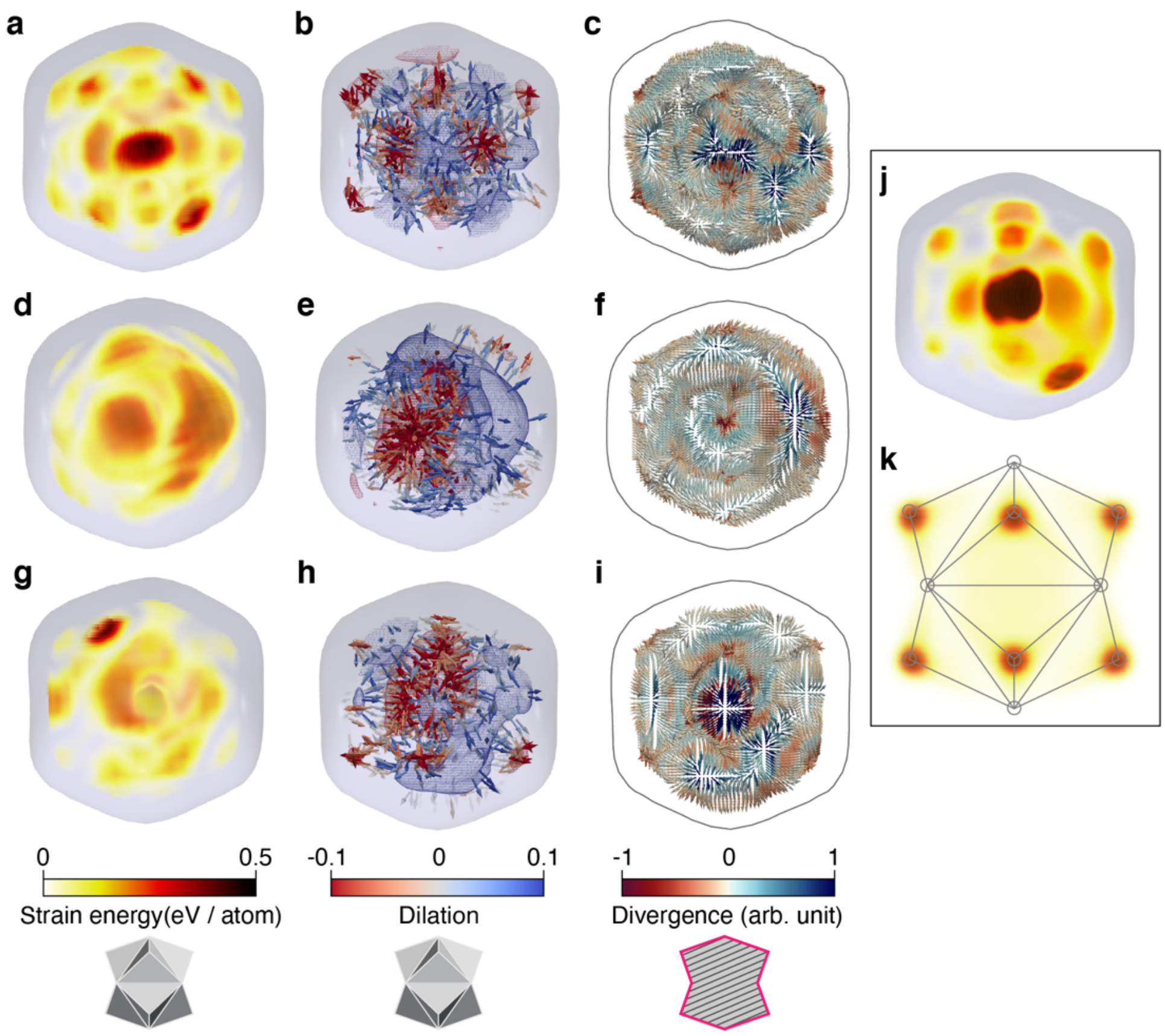

**Fig. 5| 3D map of strain energy and strain field of the NPs. a,** 3D distribution of strain energy accumulated on the type-I NP is shown. Higher strain energy is found at the position corresponding to the vertices of the trisoctahedron. **b,** The strain field in 3D is shown using arrows indicating the direction for higher strain energy. The strain energy distribution is overlaid using contour map with the regions of local expansion in blue and compression in red colors. **c,** Strain visualized using color-scaled arrows exhibits locally contracted and relaxed regions at the central plane of NP. Similar plots of 3D strain energy distribution (**d, g**), 3D strain field (**e, h**) and detailed strain field on a plane **(f, i)** are obtained for type-II and type-III NPs, respectively. **j,** Strain energy distribution of the type-I NC calculated from the MD. **k,** Strain energy distribution expected for the ideal TOH Au NC obtained from MD simulations is compared.

## Methods

**Au@$TiO_2$ core-shell nanoparticles.** To prepare the trisoctahedral (TOH) Au nanocrystals (NCs), Au NC seeds were grown in two steps. In a typical synthesis of Au NC seeds, 0.5 mL of $HAuCl_4$ (5 mM) and 0.6 mL of ice-cold $NaBH_4$ (10 mM) were sequentially injected into a 10 mL aqueous solution of cetyltrimethylammonium chloride (CTAC, 100 mM) with vigorous stirring for 1 min. The resultant Au NC solution was incubated at room temperature for 5 hours and used as a seed solution for the larger size of Au NCs. In the next step, 2 mL of $HAuCl_4$ (5 mM), 1 mL of L-ascorbic acid (AA, 300 mM) and 0.1 mL of 10-fold diluted Au NC seed solution were sequentially added into a 20 mL aqueous solution of CTAC (100 mM), and then the mixture was stored at room temperature for overnight. The resultant Au NCs were used as a second seeds. For the synthesis of TOH Au NCs, 2 mL of $HAuCl_4$ (5 mM), 0.2 mL of second Au NCs seed solution and 1 mL of L-ascorbic acid (AA, 300 mM) were sequentially added into a 20 mL aqueous solution of CTAC (100 mM), and then the resultant solution was incubated into room temperature for overnight.

In a typical synthesis of TOH Au@$TiO_2$ core-shell NPs, 10 mL of TOH Au NC solution was centrifuged and the collected NCs were re-dispersed into 20 mL of water. Into this solution, 40 μL of 11-mercaptoundecanoic acid (MUA, 50 mM in ethanol) was added and the mixture was shaken at room temperature for 2 hrs. The MUA treated TOH Au NCs were isolated by centrifugation and re-dispersed into the mixture of 0.2 mL of water and 20 mL of ethanol. After adding 70 μL of titanium (IV) butoxide (17 vol % in ethanol) into the solution, the resultant mixture was shaken at room temperature for 12 hrs. The final product was washed with ethanol and dispersed in ethanol for analysis.

**XFEL single-pulse diffraction imaging experiments.** XFEL single-particle 3D imaging experiments have been carried out at the second hard X-ray experimental station (EH2) of the PAL-XFEL. Incident X-ray energy was tuned to 5 keV using a 100 meter-long undulator array to accept the native SASE bandwidth of $\Delta E/E \sim 5\times10^{-3}$ without employing additional monochromator. A pair of K-B focusing mirrors was installed at 5.0 m upstream of the sample interaction spot, which has focused the XFEL pulses to the size of ~ 5 μm (H, horizontal) × 7 μm (V, vertical) at full width at half maximum (FWHM). The average number of XFEL photons in each pulse is $\sim 8\times10^9$ photons $\mu m^{-2}$ at the sample position[41]. Single-pulse diffraction patterns were recorded by a one-megapixel multi-port charge-coupled-device (MPCCD) detector installed at 1.6 m downstream of the sample.

**Fixed target, single particle three-dimensional imaging.** NPs were dispersed in deionized water solution with a concentration to increase the single-particle hit event[24]. Droplets from this solution are then sprayed onto the $Si_3N_4$ membranes that were plasma cleaned to enhance their wettability. Each membrane holds a 36-by-36 array of 100 nm thin $Si_3N_4$ windows with 200 μm x 200 μm area for each window, custom designed for single-pulse diffraction experiments. We collected single-pulse two-dimensional (2D) diffraction patterns of specimens at random orientations by scanning the focused XFEL radiation across the membrane window. Specimens that were illuminated by single intense XFEL pulses were completely destroyed leaving behind a hole on the window. This destruction precludes the possibility of hitting the same sample in multiple times. Spherical morphology of the shell facilitates its deposition on a flat membrane without preferred orientation, which allows adequate sampling of different 2D projections for the 3D imaging.

The average number of particles on each $Si_3N_4$ membrane was ~ $5\times10^7$ to have the number density of 0.08 particles $\mu m^{-2}$. We exposed 32,400 XFEL pulses on a membrane to acquire ~ 320 copies of high-quality, single-particle diffraction patterns used for 3D reconstructions. This led to the sample demands for one high-quality single-particle diffraction pattern of $1.6\times10^5$ particles.

**Automatic screening of single-particle diffraction patterns.** We processed each diffraction pattern by subtracting the background noises and aligning the diffraction center, and numerically binning the data over 3-by-3 pixels into one. Binned diffraction patterns were cropped to 299-by-299 array, originally 897-by-897, and put into pretrained Xception deep neural network[26]. The predictions on 1000 classes of images were calculated. By examining the most likely class of randomly picked images, we selected most abundant 10 classes that were related to one of types: no-, single-, multi-particle diffraction patterns and irrelevant patterns from broken substrate or aggregated particle.

In order to classify patterns, we constructed feature vectors with the predictions on these 10 classes and extracted the principal components. From the distribution of the test dataset on the multi-dimensional feature space, we projected the data along the two principal components ($PC_1$ & $PC_2$). Then the hit rates of each diffraction type were estimated from 100 randomly sampled patterns at every grid. The final 1,176 diffraction patterns were selected after visual inspection of the data on grids where the single hit rate was larger than 1% in order to filter out incorrectly synthesized nanoparticles such as bipyramidal particles, and contaminants from synthesis solution, dust on substrate, etc.

**Assembling a three-dimensional diffraction pattern using EMC.** The EMC algorithm was modified to reconstruct multiple three-dimensional (3D) intensity volumes from 1,176 single-particle diffraction patterns. Overall, this multiple-model EMC (mmEMC) algorithm inferred the orientation distribution (with 50,100 quaternion samples), local photon fluence, and model probability of each pattern. To increase the orientation coverage of the patterns, we fractionated each pattern into 100 random sub-patterns that added back to the original pattern (Supplementary Methods for details). The octahedral symmetry of any reconstructed 3D volume, even when this symmetry is not enforced during the reconstruction, is evident in the preferred orientation of fractionated patterns that show strong affinity to the volume. From extensive testing, we found the 4-model mmEMC reconstruction attempts gave the most reproducible 3D volumes (Supplementary Methods). To ensure minimum overlap of pattern occupancy between these reconstructed models, we used the four models reconstructed via 4-model mmEMC that produced the most reproducible volume (Fig. 1d).

**Angular correlation for symmetry identification** The angular correlation map was obtained by calculating the cross-correlation between the posterior probability distribution of the 2D diffraction patterns as a function of relative rotation Ω to the 3D model. The rotational cross-correlation $\xi_{rot}$ is,

$$\xi_{rot}(\Omega'|K,W) = \int_{\Omega \in \mathrm{SO}(3)} p(\Omega|K,W)p(\Omega \cdot \Omega'|K,W)\mathrm{d}\Omega,$$

where Ω, K, W is the 3D rotation, 2D diffraction pattern, 3D model, respectively, and $\Omega'$ is the relative 3D rotation in cross-correlation. Angular correlation $\xi_{ang}(\theta|K,W)$ is then obtained by summing the correlation values of 3D rotations with the same angle of rotation. An $n$-fold rotational symmetry results in high cross-correlation at rotation angles of $2\pi/n$, and these peaks are prominent for the majority of patterns for each 3D volume. Because the symmetry peaks appear commonly along whole patterns, we averaged the angular correlation of whole patterns to obtain the angular correlation map (Fig. 3e).

**Numerical phase retrievals for 3D image reconstruction.** We have carried out numerical phase retrieval to acquire 3D density maps using the assembled 3D diffraction volumes. Recently introduced GPS-F algorithm was employed with its improved performance showing better convergence for the single-pulse XFEL diffraction patterns with significant shot noise[42]. For each 3D diffraction pattern, 400 independent reconstructions have been carried out starting from random seeds (Supplementary Table 1). Each phase retrieval ran for 1,000 iterations, and the best 200 images with the lowest K-space errors have been averaged to represent the image of the diffraction pattern (Supplementary Methods for more details).

**Coherent diffraction imaging using synchrotron X-rays.** The coherent diffraction imaging experiments were carried out at 9C Coherent X-ray Scattering (CXS) beamline of the Pohang Light Source. Incident X-ray energy was tuned to 5.4 keV using a Si double-crystal monochromator. X-ray absorption by the specimen at this energy is slightly smaller (~ 3 %) to the 5keV XFEL experiments. A flat mirror was installed to filter out higher harmonic components of the incident X-ray radiations. Before the specimen, the K-B mirror was installed to focus the X-ray beam to have focused beam size of 9 μm (H) × 12 μm (V) formed at 3.0 m downstream of the second K-B mirror. The Au@$TiO_2$ NP specimens, from the same sample growth, were mounted on a $Si_3N_4$ membrane same as the XFEL experiments. The photon flux density of the focused beam was $3.5\times10^8$ photons $\mu m^{-2}$ $s^{-1}$. Diffraction patterns were collected using the Timepix detector, which is the photon counting detector with the peak dynamic range of 11,810 photons.

**Gaussian curvature estimation and identification of facet indices** The Gaussian curvature ($K$) of the surface is defined as a product of two principal curvatures, $\kappa_1$ and $\kappa_2$. For a perfect sphere with radius, $R$, its Gaussian curvature is $R^{-2}$. In order to calculate principal curvature of the TOH Au core, we first determined the surface of the Au core with the density threshold of 50 % of average core density. Then a quadric surface was locally approximated on each vertex of the Au surface and their nearest neighbors with '*Curvature*' function in Avizo software (ThermoFisher Scientific). Four attempts were averaged to reduce fluctuations introduced during the numerical approximation of the surface.

From the defined Au surface, 8 pyramidal vertices and 6 octahedral vertices were identified displaying the local maxima in the surface curvature, and the distances from the center were calculated (Supplementary Fig. 6). The vectors connecting the local maxima positions were calculated to determine {111} planes and three high-index-facetted $\{hh\bar{l}\}$ planes using vectors connecting one pyramidal vertex and adjacent three octahedral vertices. Miller indices were calculated from the angle between the high-index facet planes $\{hh\bar{l}\}$ and {100}. More details can be found in the Supplementary Methods.

**Atomic structure model from molecular dynamics simulations** The atomic model of the Au core NC was built as follows. Initial model was generated using Au atoms in face-centered cubic lattice. The atomic model complying with the obtained 3D structure was constructed by numerically patching small volumes (2.44 nm x 2.44 nm x 2.44 nm) of homogeneous density to have the overall 3D density consistent with the 3D density map obtained from experiments. The assembled NC structure was then optimized using molecular dynamics as implemented in LAMMPS[43]. The relaxation of the assembled NC structure was achieved in multiple steps including heating of the boundary atoms and relaxation of the entire NC. Atomic structure model for the ideal TOH Au NC was established using ideal, non-distorted, face-centered cubic Au of {331} facets with homogeneous density. The optimization of the ideal structure involved the heating and relaxation of the entire NC structure only. The equation of motion was integrated using the velocity-Verlet method with the time step of 0.5 fs. The strain energy per atom for the ideal TOH Au NC structure was obtained by calculating atomic strains. The strain energy of the model structure of type-I NC was obtained from the local volume dilation, to prevent over-interpretation of the atomic model constructed artificially. Details of the structure relaxation procedure and strain energy calculation are found in the Supplementary Methods.

**Data availability**

All relevant data are available from the corresponding authors.

**Acknowledgments**

C.S. appreciate helpful comments from J.S.Kim. The experiments at the PAL-XFEL and the Pohang Light Source (PLS) in Pohang Accelerator Laboratory (PAL) were approved by Korean Synchrotron Users Association. Funding: This work was supported by the National Research Foundation (NRF) of Korea (Grant No. 2015R1A5A1009962, 2019R1A2B5B03070059, 2015R1A3A2033469). Y.I. was

supported by NRF of Korea (No. 2018R1D1A1B07040727). S.Z. and N.D.L. were supported by NUS Startup grant.

**Author contributions**

C.S. conceived the project. D.C. analyzed the data and performed 3D image reconstructions. Z.S. and N.D.L. carried out mmEMC calculation. Y.I. performed the MD simulations. D.C., C.J., D.S., D.-W.N., H.L., J. S., J.H., S.-S.K., S.-Yu.L, S.-Yo.L., K.S. K, D.-Y.N. and C.S. have contributed to the experiments. D.H.W. and S.W.H. prepared the core-shell nanocrystals and carried out EM characterizations. D.C., Z.S., Y.I., D.H.W., S.W.H., N.D.L. and C.S. wrote the manuscript with the inputs from all authors.

**Competing interests**

The authors declare no competing interests.

**Additional information**

**Supplementary information** is available for this paper

**Correspondence and request for materials** should be addressed to C.S. or N.D.L.

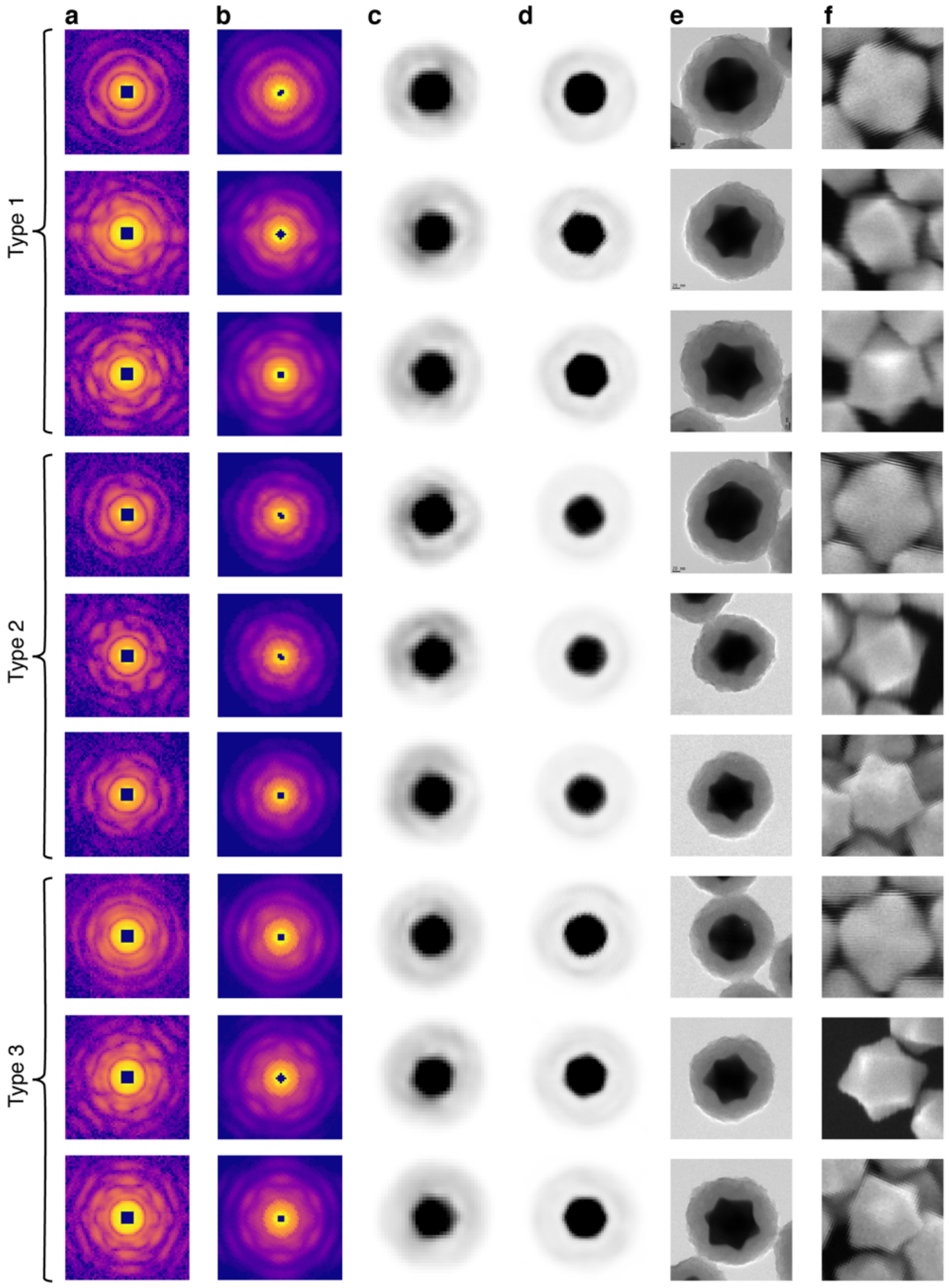

**Extended Data Fig. 1| Comparison of the 3D structure with 2D projection images and electron microscope images. a,** XFEL single-pulse diffraction patterns. **b**, 2D sections extracted at the same orientation from the reconstructed 3D diffraction volume. **c,** Reconstructed images of 2D single-pulse patterns. **d,** Projection images from the 3D reconstructions. **e,** TEM images of the NPs with similar morphology. **f,** SEM images of the TOH Au NC with similar morphology. The results are compared for type-I, type-II and type-III structures at different orientations.

Supplementary Information

# High-throughput ensemble characterization of individual core-shell nanoparticles with quantitative 3D density maps from XFEL single-particle imaging

Do Hyung Cho, Zhou Shen, Yungok Ihm, Dae Han Wi, Chulho Jung, Daewoong Nam, Sangsoo Kim, Sang-Youn Park, Kyung Sook Kim, Daeho Sung, Heemin Lee, Jae-Yong Shin, Junha Hwang, Sung-Yun Lee, Su Yong Lee, Sang Woo Han, Do Young Noh, N. Duane Loh, Changyong Song

## List of Supplementary Information

# Supplementary Methods

## Sections

1. Multi-model EMC
2. 3D phase retrieval
3. Identification of facet indices
4. Elastic strain energy calculations
5. Relaxation of the assembled structure using MD

### 1. Multi-model EMC (mmEMC)

#### A. Introduction

To resolve the structural heterogeneity between different illuminated particles, we devise the multi-model EMC (mmEMC) algorithm designed to extract structural classes from the dataset without human supervision. As its name suggests, mmEMC can reconstruct more than one 3D intensity model from a given set of diffraction patterns. Patterns (i.e. frames) cluster towards the reconstructed models from which they are likely to arise.

For situations where the particles are fairly heterogeneous, such that the structure of most particles are detectably different at the captured resolution, mmEMC can serve as a structural classifier. Once such a dataset is classified, we can then extract and retrieve the phases of the most reproducible reconstructed models.

The mmEMC is a generalization of the original EMC algorithm (implemented in the package Dragonfly), whereby each pattern's latent variables is extended from its 3D orientation to include its structural model[1]. Following Dragonfly's notation, the original latent variable (orientation) index $r$ is now replaced by a tuple, ($m$, $r$) (or just $mr$), where $m$ is the index of model. The fundamental variable in Eq. (2) in Dragonfly, $W_{rt}$, is rewritten as $W_{mrt}$, which refers to the tomogram of model $m$ at the Ewald-sphere orientation $r$. We can write down mmEMC's iterative update of the model ($W \rightarrow W'$) and latent fluence factor ($\varphi_d \rightarrow \varphi_d'$) as (details in section C),

$$W_{mrt}' = \frac{\sum_d P_{dmr} K_{dt}}{\sum_d P_{dmr} \varphi_d}, \qquad (1)$$

$$\varphi_d' = \frac{\sum_t K_{dt}}{\sum_{mrt} P_{dmr} W_{mrt}}, \qquad (2)$$

## B. Fractionating a single frame into subframes

The goal of recovering the 3D diffraction volume from many diffraction frames, in broad terms, is equivalent to retrieving the orientations for each frame. Although we used 50,100 quaternions sampling this orientation space here, there are only 1,176 frames in the dataset. At this angular sampling, because the frames in this dataset have relatively high signal-to-noise ratio, each frame would almost certainly be locked into one orientation. Consequently, the coverage of the orientations is less than 3%, which is too low for EMC to converge stably.

To improve the data coverage of the orientation space we split one frame into $N$ subframes, whose total photon counts per pixel adds back up to those in the original frame. This is essentially a regularization procedure that increases the orientation uncertainty of each pattern, which in turn leads to greater orientation coverage. For this split, first the photons in each of the pixels are randomly and independently distributed across $N$ partitions. We then combine a random partition from each pixel (without replacement) together to create a single subframe. We empirically choose the subframe number $N$ by increasing the number until the reconstruction becomes stable, which ended up $N$=100. We note that the choice of $N$ is not crucial because all conditional probabilities of subframes are merged first.

## C. Updating fluence factors in mmEMC with frame fractionation

We now consider another latent, unmeasured quantity of the mmEMC reconstruction: local pulse intensity. Each particle encounters different regions of different X-ray pulses, and consequently, different particles experience different effective local photon fluence factor, $\varphi(K)$. Hence, the differences in fluence factors associated with different photon patterns must be corrected before they can be merged into a single three-dimensional (3D) diffraction volume. Equivalently, each pattern has be downscaled by $\varphi(K)$ to before merging. This rescaling strategy is equivalent to the one in Dragonfly, and similar to those in previous EMC reconstructions[2,3]. However, this strategy must be modified to accommodate the multiple model and sub-frame fractionation approach used here.

Without loss of any structural information, any 3D model $m$ can be rescaled by an overall multiplicative constant. For simplicity, we can always rescale any 3D model such that the average photon fluence factors $\varphi(K)$ of its likely frames is unity. This $m$-dependent rescaling factor, $\bar{\varphi}_m$, is implemented as,

$$\bar{\varphi}_m \equiv \sum_K p(K \mid W_m)\varphi(K) = 1, \qquad (3)$$

where $p(K|W_m)$ is the likelihood that frame $K$ was generated from 3D model $W_m$. Equivalently, the 3D model $W_m$ must be multiplied by factor $\bar{\varphi}_m$ to maximize its likelihood of generating the set of frames $\{K\}$. Overall, this last constraint helps accommodate different frames into the multiple models reconstructed by mmEMC.

However, because a frame $K$ might have high likelihoods for multiple models, each of which has a different rescaling factor $\bar{\varphi}_m$, there must be a constraint that ensures that each frame concurs with these different models. To enforce this concurrence, we rewrite the fluence factor update step of our mmEMC iteration ($\varphi(K) \rightarrow \varphi'(K)$) as

$$\varphi'(K) = \frac{\varphi(K)}{\sum_m \bar{\varphi}_m p(W_m \mid K)}, \tag{4}$$

which now re-weights the fluence factor a frame $K$ by the scaling factor of the models from which it is likely to arise.

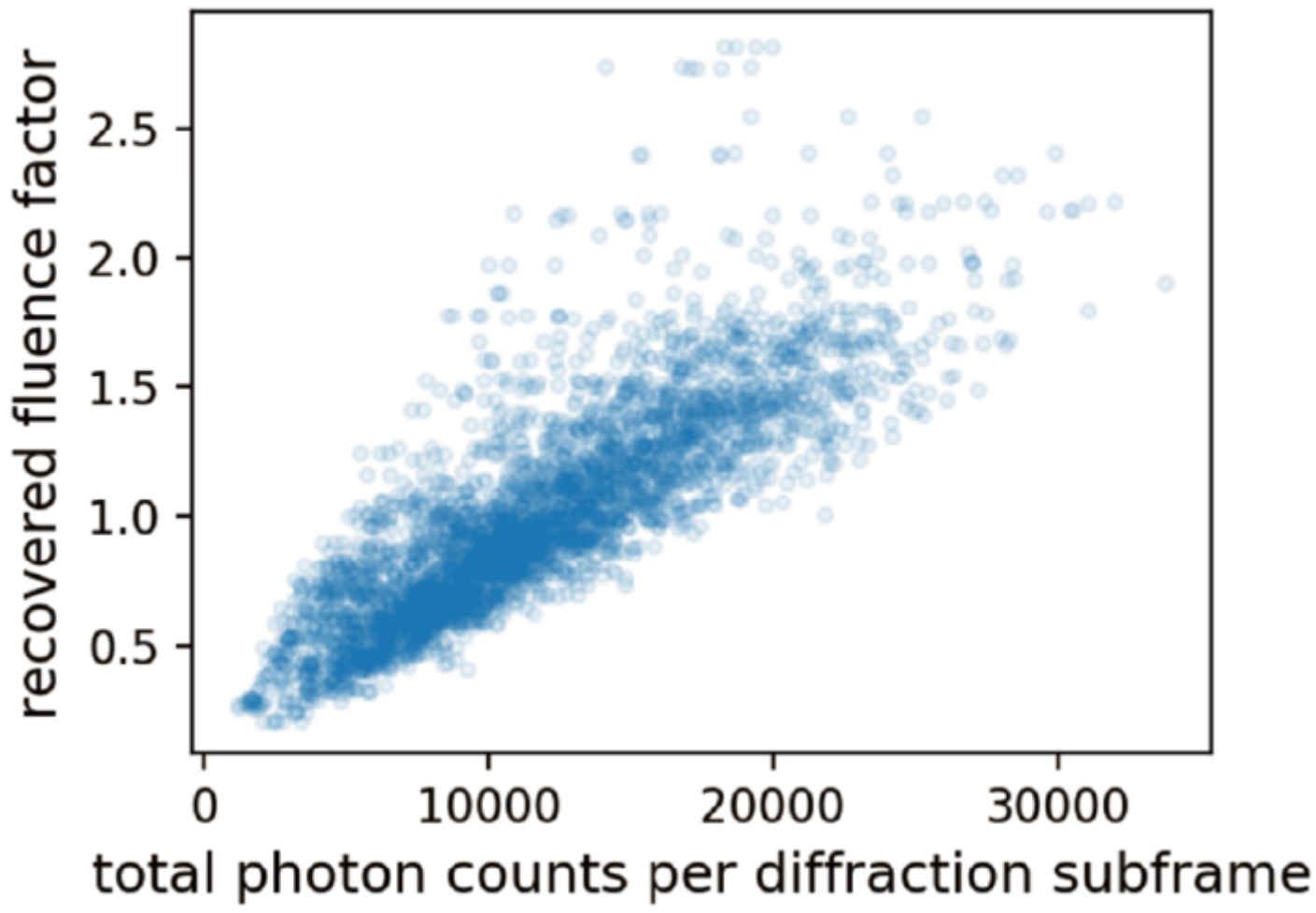

**Supplementary Fig. 1| Correlation between total photon counts and fluence factor**. This figure shows a positive correlation between the local photon fluence factor recovered by mmEMC versus the total photon counts in each diffraction subframe. The notable spread in this distribution is a result of maximizing the likelihood that pattern of photons on these diffraction subframes concur on the same set of 3D diffraction models.

Furthermore, fractionating a single diffraction frame into several subframes adds considerations about how a model's average fluence factor should be updated. Note that Eq. (2) reconstructs a different fluence factor for each subframe (Supplementary Fig. 1). Not only impose all subframes originating from a single raw frame to have the same fluence factor, we also directly constrain the average fluence factor of these subframes. The variations between the fluence factors between subframes of a single original frame is part of the regularization of orientation space described in section B.

### D. Extracting particle rotation symmetries from the posterior distributions over the latent space of orientations and models

An exhaustive search for all the possible rotation symmetries of a 3D diffraction volume can be computationally expensive. One 'brute-force' approach is to compare the 3D volume with a rotated version of itself around a sufficiently large number of candidate axes of rotational symmetry. A simple implementation of this scheme demands computation time that scales like the diameter of the particle to the fifth or sixth power.

Instead, we take a different approach that exploits the sparsity of the posterior distribution of each photon frame that is likely to arise from any particular 3D model. On average, the computational demands here scales like the number of diffraction frames multiplied by the square of the number of likely orientations of each pattern. For this experiment, this calculation is purportedly 100-fold faster than the 'brute-force' approach in the previous paragraph.

Suppose a specimen has perfect four-fold rotation symmetry around a certain axis $G$. Consequently, any Ewald sphere section (or their corresponding diffraction frames) of the specimen's diffraction volume should have high likelihood at the four orientations on the symmetry orbit around $G$. Within the context of an EMC reconstruction, such rotational symmetries of the specimen, $W_m$, should be reflected in the posterior distribution $p(\Omega \mid K, W_m)$ over the latent space of orientations $\Omega$ given the frames $K$ that it is likely to generate. Specifically, a high fidelity single-model reconstruction $W$ (i.e. $m$=0 only) must also show a corresponding symmetry in its posterior distribution over orientations:

$$p(\Omega|K,W) = p(\Omega_{\mathrm{G}} \cdot \Omega|K,W), \quad (5)$$

where $\Omega_G$ is the rotation operator that steps through this symmetry orbit. If subframes are used, then the left-hand side of Eq. (5) should average over the subframes that comprise each frame ($K_S \in K$),

$$p(\Omega|K, W) = \langle p(\Omega|K_S, W)\rangle_{K_S}. \quad (6)$$

Equations (5) and (6) imply that the rotational autocorrelation of the posterior $p(\Omega|K, W)$ should share the same symmetries as the rotational symmetry of a specimen. This rotational autocorrelation is denoted

$$\xi_{rot}(\Omega'|K, W) = \int_{\Omega \in \mathrm{SO}(3)} p(\Omega|K, W) p(\Omega \cdot \Omega'|K, W) \mathrm{d}\Omega, \quad (7)$$

is computed here only for values of $p(\Omega|K, W)$, which is typically sparse in $\Omega$, that exceed some threshold.

In this work, we are only concerned with the *order* of the specimen's rotation symmetry (e.g. 2-fold symmetry, 4-fold, etc.) rather than its subgroup's conjugate representation in SO(3). Because the axis-angle (i.e. $\Omega$–$\theta$) representation of an orientation is a class function, the order of a particle's rotation symmetry can be found by mapping its axis of symmetry $\Omega$ into $\theta$. This mapping of $\xi_{rot}(\Omega|K, W)$ to $\xi_{ang}(\theta|K, W)$ is also advantageous from a signal-to-noise perspective because the total likelihood from a particular number of frames is mapped from the three-dimensional $\Omega$ space into a smaller $\theta$ space.

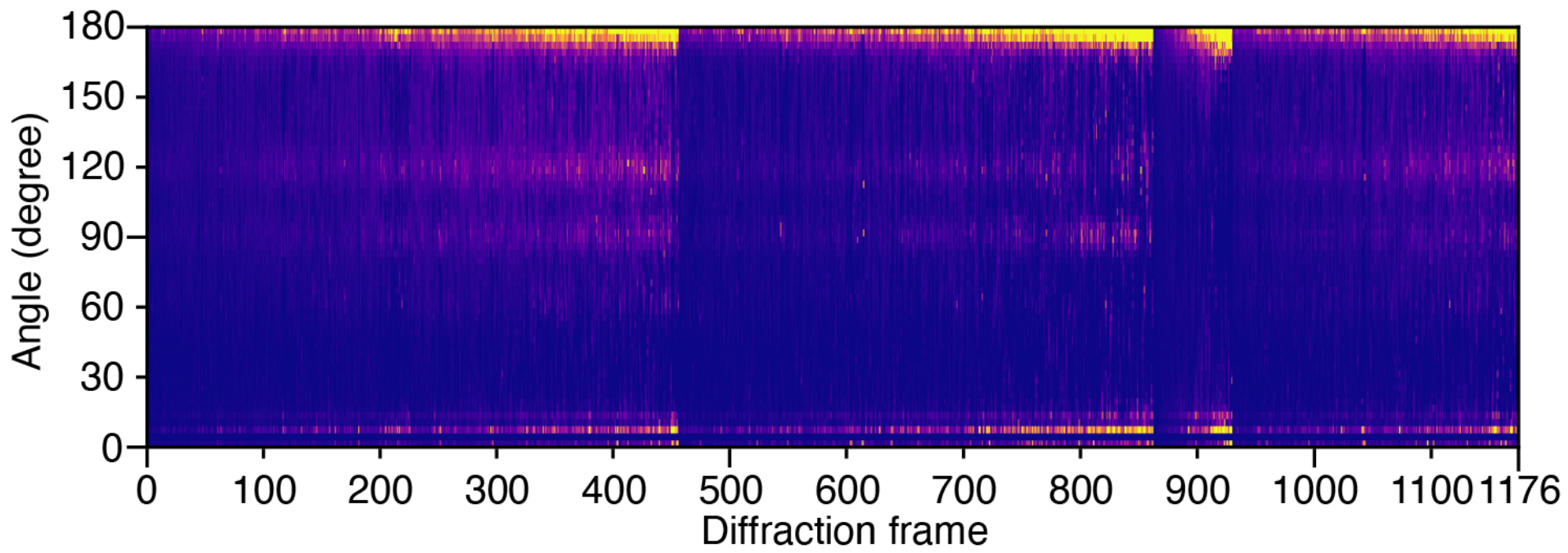

**Supplementary Fig. 2 | Posterior probability of reconstruction.** The projection of model posterior probability autocorrelation $\xi_{ang}(\theta|K, W)$ reveals the octahedral symmetry of the reconstructed models.

The gold cores of the nanoparticles in this experiment are expected to have octahedral symmetry. This symmetry group comprises six two-fold symmetry subgroups, three four-fold symmetry subgroups and four three-fold symmetry subgroups: $\theta(\Omega_G)$=0°, 90°, 120°, 180°. As shown in the Supplementary Fig. 2, most of reconstructed models have these four peaks.

## E. Comparing structural models in mmEMC

In this section, we wish to establish if a particular reconstructed 3D intensity model is readily reproducible given a set of diffraction frames. Consider the mmEMC reconstruction attempt $A$ that comprises a number of 3D intensity models, indexed by $m$. The idea here is to use these diffraction frames to vote on the reproducibility of 3D models. If a particular model $W_{Am}$ is reproducible, it should recur in at least another independent mmEMC reconstruction $B$. A particular 3D model would show a high likelihood of generating a particular subset of diffraction frames. If this same subset of frames again shows a high likelihood for another 3D model, then these two 3D models should presumably be fairly similar.

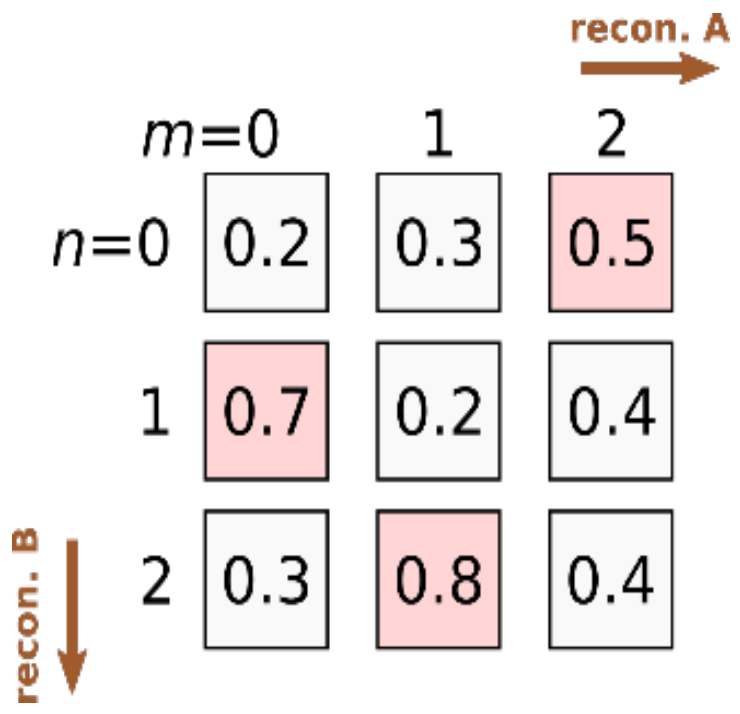

**Supplementary Fig. 3| Comparing models in between different mmEMC reconstructions.** Here is an example of the matrix elements of similarity measure Eq. (8), which is the dot product of their posterior probabilities for models within either reconstruction $A$ (indexed by $m$) or $B$ (indexed by $n$). Matching the most similar models $W_{Am}$ to $W_{Bn}$ is equivalent to solving the linear assignment problem.

However, the model labels between reconstructions $A$ and $B$ are random and unmatched. To match these labels up, we compute the similarity between the model posterior probabilities of them,

$$S_{mn} = \langle p(m\,|K, W_A)\,p(n|K, W_B)\rangle_K, \quad (8)$$

where $W_A$ and $W_B$ are the reconstruction $A$ and $B$, and $K$ are all the measured diffraction frames used to reconstruct $A$ and $B$. Comparing against the $m$th model in $A$, the most similar model in $B$ should have the largest similarity $S_{mn}$. A matching scheme should consider all matched pairs, which means the optimizing target is the summary of $S_{mn}$ of each pair of $(m, n)$ in the scheme. This is classical maximum weight matching problem. Picking $A$'s model labels as reference, the models in the remaining reconstructions can be matched up using $S_{mn}$ (see caption in Supplementary Fig. 3).

Our goal here is to find the least *populated* and most reproducible model among all the reconstructions. For the first criterion, frames belonging to structurally distinct specimens sometimes tend to aggregate together into a single model, which resembles an incoherent average of different scattering volumes. To reduce this undesirable averaging, we preferentially seek models that show high likelihoods for smaller number of frames. With just this single criterion, however, we risk selecting models that were overfitted too few frames. Because these overfitted models tend to be random and irreproducible, we use a second reproducibility criterion to weed them out. To do so, we find the two independently reconstructed models that are most mutually similar.

Assuming the model indices in $A$ and $B$ are matched according to the similarity metric in Eq. (8), the similarity between models $W_{Am}$ and $W_{Bm}$ is reflected in their frame posterior probabilities, $p(m|K, W_A)$ and $p(m|K, W_B)$. For this comparison, we constructed a posterior matrix for model $W_{Am}$:

$$\mathcal{P}_{Am} = \{\,\mathbf{P}(K, W_A) \mid K \in \mathcal{K}_{Am}\,\}$$

where the model pattern set

$$\mathcal{K}_{Am} = \{K \mid \arg\max_n p(n \mid K, W_A) = m\};$$

the posterior vector

$$\mathbf{P}(K, W_A) = \big(p(0|K, W_A), p(1|K, W_A), \dots, p(|A| - 1|K, W_A)\big),$$

where $|A|$ is the number of models used in reconstruction $A$.

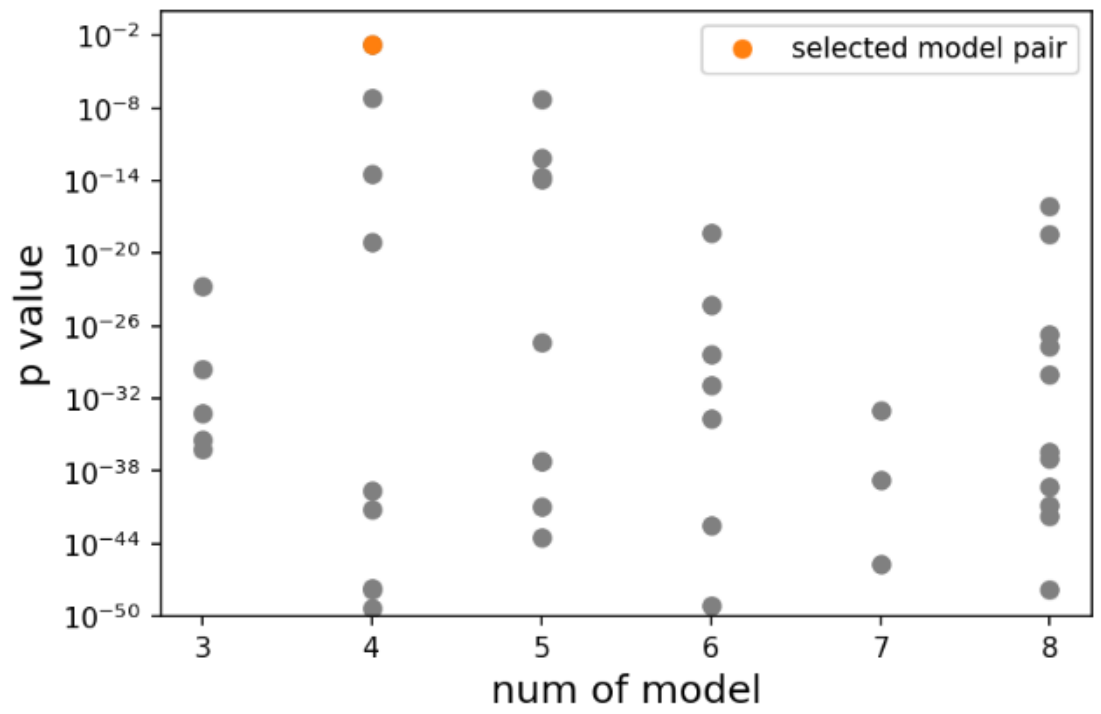

**Supplementary Fig. 4| Determining proper number of models by statistical test.** The p-value calculated with Kruskal-Wallis test for all pairs of matched models between the reconstructions with same number of models. The orange dot is the selected model pair with highest p-value.

We used the Kruskal-Wallis non-parametric test[4] to decide if two posterior matrices $\mathcal{P}_{Am}, \mathcal{P}_{Bm}$ are drawn from the same distribution (i.e. null hypothesis). If the null hypothesis cannot be rejected, then $W_{Am}$ and $W_{Bm}$ are considered the same model. To obtain the canonical sorted ranks for this test, we project the posterior vectors in $\mathcal{P}_{Am}$ and $\mathcal{P}_{Bm}$ onto a randomly picked posterior vector $\mathbf{v}_r$ whose length equals $|A|$. This projection reduces the posterior matrices $\mathcal{P}_{Am}$ and $\mathcal{P}_{Bm}$ each into one-dimensional list of real numbers. The rank of these real numbers within union of both lists are then used in the Kruskal-Wallis test. To speed up this computation between two reconstructions, we only evaluate this test between matched pair of models $(Am, Bm)$. To reduce the bias from any random vector, we picked the smallest $p$-value amongst a thousand random vectors $\mathbf{v}_r$, where the corresponding $\mathbf{v}_r$ is the most separable direction of $\mathcal{P}_{Am}$ and $\mathcal{P}_{Bm}$ (Supplementary Fig. 4).

In all matched pairs between all reconstructions each with a specified number of models, a 2-model reconstruction pair has the highest $p$ value. This pair, however, was populated by more than 90% of the frames, which almost decays to a single model reconstruction pair. The next highest $p$-value belonged to a 4-model reconstruction pair whose frame occupancy was about 50%. Considering the two criteria for the least *populated* and most reproducible model pairs, we eventually selected our model populated by 41% frames from the latter pair. This model was the type-I model in the main text (Fig. 1).

From the Supplementary Fig. 4, we also observe that 4-model mmEMC produced the most reproducible reconstructions (compared with say 5-model or 6-model mmEMC). Types I-IV in the main text comes from the specific 4-model mmEMC reconstruction that produced the type-I volume.

## 2. 3D phase retrieval

**Supplementary Table 1. Phase retrieval parameters**. $\sigma$ is the relaxation parameter and $s$, $t$ are the step sizes of the proximal mappers $\text{prox}_{tg_\sigma}$ and $\text{prox}_{sf_G^*}$. $\alpha$ is the root-mean-square of Gaussian filter for smoothing in Fourier-space.

| **Iteration #** | $\sigma$ | $s$ | $t$ | $\alpha$ **(nm)** | **Support** |
|---|---|---|---|---|---|
| 1~100 | 0.01 | 0.9 | 1 | 1730 | initial support: sphere<br>(R=190 nm,~40% larger than final one) |
| 101~200 | 0.01 | 0.9 | 1 | 1550 | |
| 201~300 | 0.01 | 0.9 | 1 | 1380 | At 300, shrinkwrap (kernel size: 55 nm, cutoff 3%) |
| 301~400 | 0.01 | 0.9 | 1 | 1210 | |
| 401~500 | 0.1 | 0.9 | 1 | 1040 | |
| 501~600 | 0.1 | 0.9 | 1 | 863 | |
| 601~700 | 0.1 | 0.9 | 1 | 690 | At 700, shrinkwrap (kernel size: 41 nm, cutoff 3%) |
| 701~800 | 0.1 | 0.9 | 1 | 518 | |
| 801~900 | 0.1 | 0.9 | 1 | 345 | |
| 901~1000 | 0.1 | 0.9 | 1 | 345 | |

Assembled 3D diffraction volumes from the mmEMC were phase retrieved to obtain 3D images of the specimens following the coherent diffraction imaging method. We implemented 3D version of the generalized proximal smoothing (GPS) with adjustments parameters (Supplementary Table 1, ref [5]). Same smoothing filter introduced in oversampling smoothness was used, but its size was maintained to be larger than the object size[6]. Initial support was chosen 40 % larger size than estimated image size, which was updated twice during the iterations by shrinkwrap method with parameters in Supplementary Table 1 (ref [7]).

The error function for reconstruction was selected as Fourier-domain error $e_{\mathrm{F}} = \frac{\sum||\mathrm{FT}\{u\}|-b|}{\sum b}$, where $u$ is 3D image, and $b$ is the diffraction amplitude. For each 3D diffraction volume, we attempted 400 reconstructions. The final image of the diffraction volume was made by averaging 200 reconstructions with lower values of the error. The error values of the final images were 0.11 (type I), 0.14 (type II), 0.12 (type III).

The 3D diffraction volume reconstructed from the mmEMC displayed lower contrast than 2D diffraction patterns. It was influenced by the down-sampling of diffraction patterns that was necessary for sufficient orientation coverage and background noises[8]. These resulted in smoothed interference fringe with higher values at the destructive interference positions. We compensated this by reading only 10% of such overestimated values at destructive interference positions, i.e. deep positions of the fringe oscillation. The fractions of those compensated data points were kept as small as possible: 1.1, 1.0, and 0.8 % of the total data points for type 1, 2, and 3, respectively.

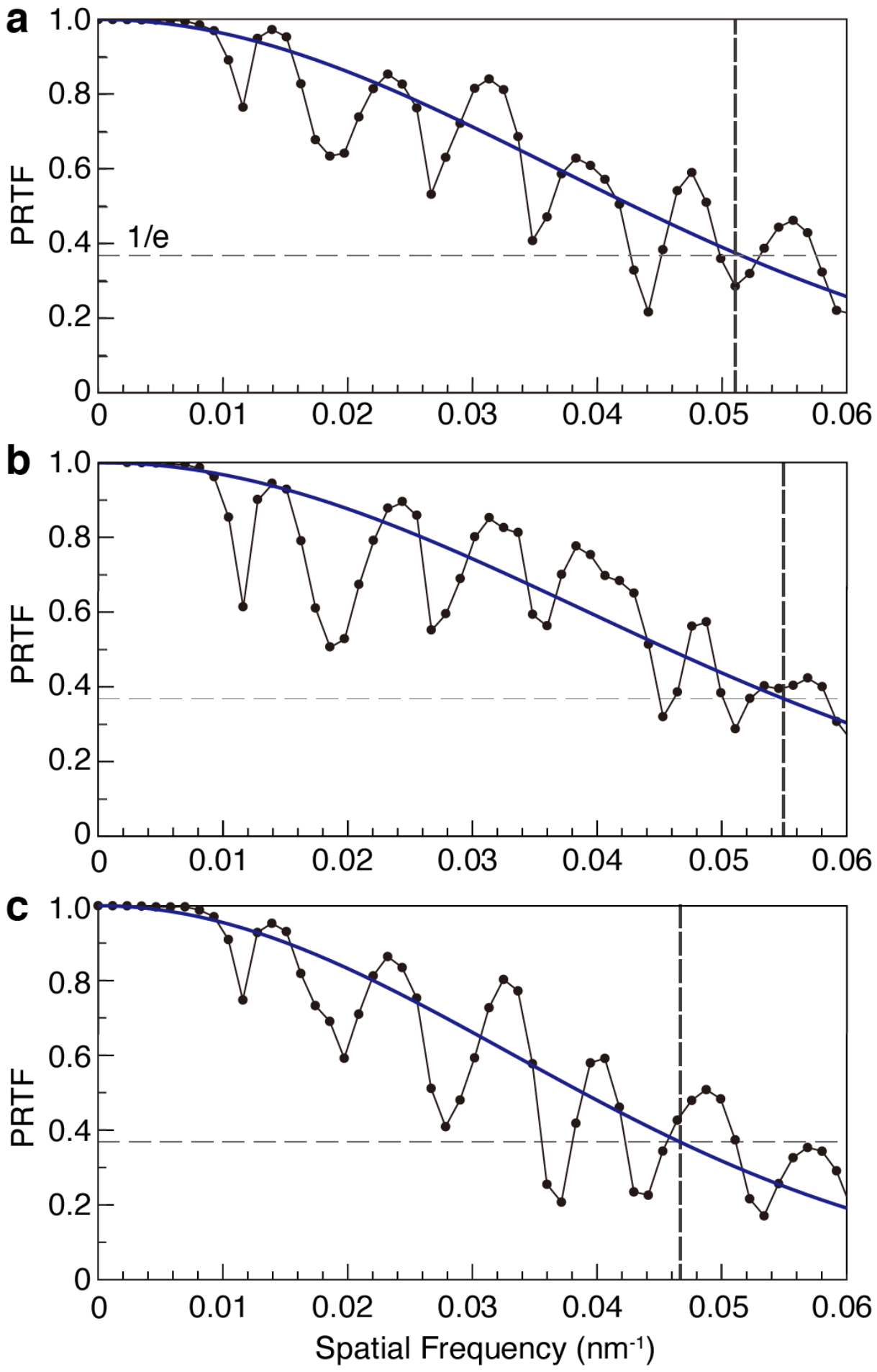

**Supplementary Fig. 5| Estimation of 3D image resolution using the phase retrieval transfer function.** 3D phase retrieval transfer functions were calculated for the 3D structures (type-I, type-II and type-III NP) reconstructed from the mmEMC assembled 3D diffraction volumes. For all the three types of 3D structures, good 3D reconstructions were obtained up to the spatial frequency of ~0.05 $nm^{-1}$ (0.051, 0.055, and 0.046 for type-I,-II and -III NPs, respectively) to result in the 3D image resolutions of ~ 20 nm.

### 3. Identification of facet indices

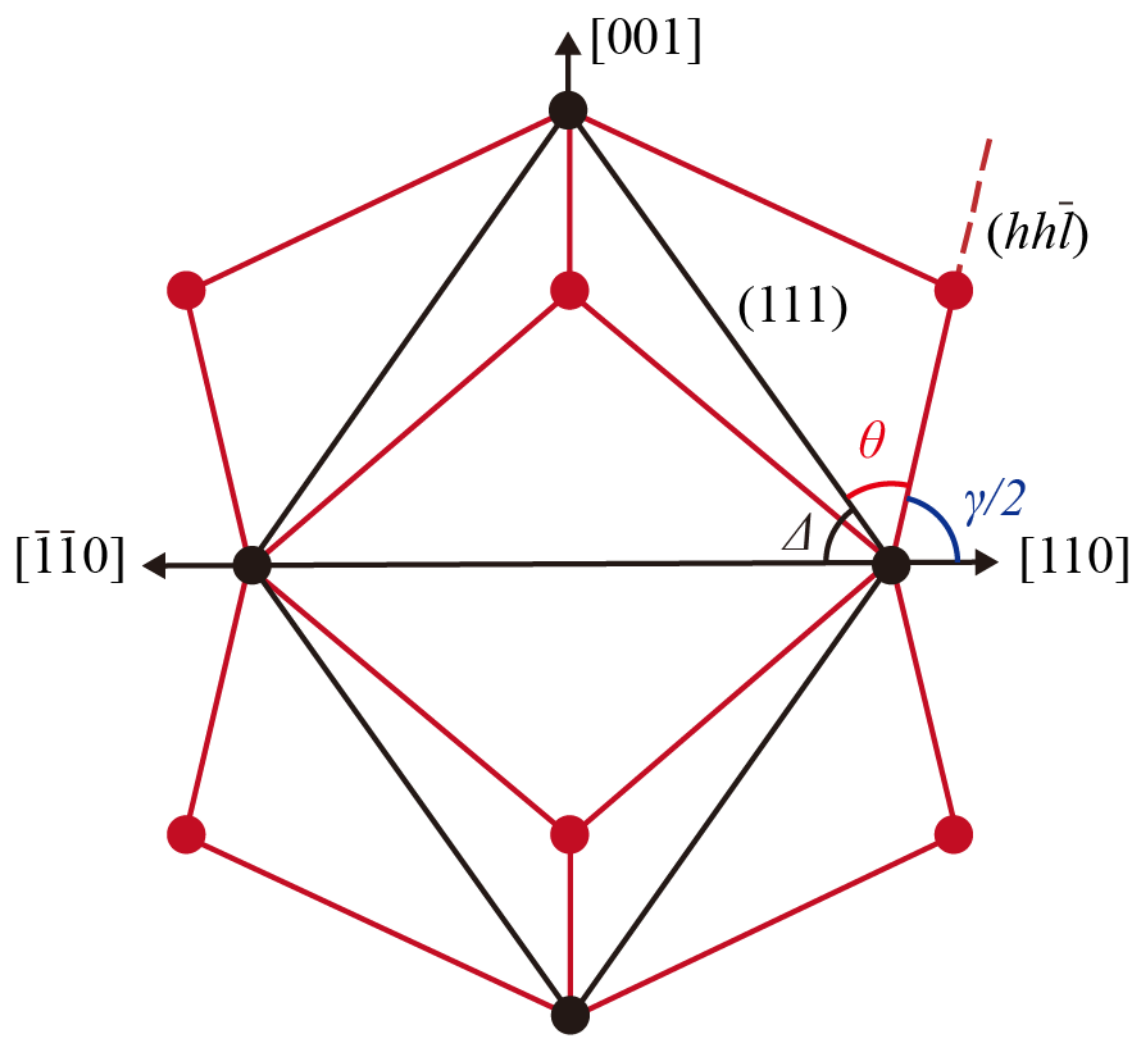

**Supplementary Fig. 6| Schematics of the facet identification.** TOH NC structure viewed along the $[1\bar{1}0]$ direction and the notations of angles described in the text.

The concave TOH structure has high-index facets $\{hh\bar{l}\}$ ($h > l >=1$). The ratio of each index $h/l$ was obtained by calculating the angle ($\gamma/2$ in the Supplementary Fig. 6) with $h/l = \tan(\gamma/2)/\sqrt{2}$; the line with slope $\tan\left(\frac{\gamma}{2}\right)$ intercepts at ($1/\sqrt{2}h$, 0) and (0, -1/$l$). We calculated the angle as $\gamma/2$=180°-$\theta$-$\Delta$. Here, the angle $\Delta$ is constant (=54.7356°), assuming the octahedron face is exactly (111) facet. The elevation angle, $\theta$, of the high-index facet can readily be calculated from the plane determined by the coordinates of vertices.

Finally, we determined the indices ($h$, $l$) that provided the lowest values of the error defined as $\frac{\theta-\theta_{h/l}}{\theta_{h/l}}$, where $\theta_{h/l} = 180° - \Delta - \arctan\left(\frac{h}{\sqrt{2}l}\right)$. For example, $h/l$ = 3.43 and our choice is (772) and its error is 0.51%. The index value was kept smaller than 7. We compared the $h/l$ calculation with TEM projection images, on 14 planes in total from 4 particles. The average $h/l$ was 3.07, which agreed well with our result.

## 4. Elastic strain energy calculation

The elastic strain energy of the TOH Au NC was approximated considering local volume expansion and contraction. One can calculate the strain energy density $u$ from the volume dilation $\delta$ by following equation,

$$u = \frac{1}{2} B\delta^2,$$

where $B$ is bulk modulus. The dilation was obtained from the density distribution $\rho(\mathrm{x, y, z})$ of the reconstructed 3D image by,

$$\delta \triangleq \frac{V - V_0}{V_0} = \frac{\rho_0}{\rho(x, y, z)} - 1$$

where $\rho_0$ is the nominal mass density of the Au crystal, $V_0$ is the unstrained volume, and $V$ is the strained volume to cause local density variation. The above relation can be derived from $\frac{\rho}{\rho_0} = \frac{V_0}{V}$. We have quantified the density obtained from the experiments with its average value equal to the nominal density.

The elastic strain energy per atom, $U$, of the reconstructed electron density (Fig. 5a, d and g) and the MD relaxed type-I model structure (Fig. 5j) was calculated as

$$U = \frac{1}{N}\frac{1}{2} B \left(\frac{\rho}{\rho_0} - 1\right)^2 V_0,$$

where $N$, $B$, $\rho$, $\rho_0$, $V_0$ are the number atoms in the reconstructed voxel, the bulk modulus of Au (180 GPa), the reconstructed density, the nominal density of Au, and the volume of the voxel, respectively. The cohesive energy of Au is 3.8 eV[12]. The length of the lattice distortion is attained from $d_0\sqrt[3]{\delta - 1}$, where $d_0$ is the nominal lattice constant (4.07 Å) and $\delta$ is the dilation.

The elastic energy of the ideal TOH Au NC was calculated as

$$U = \frac{1}{N}\frac{1}{2} B \sum_{i=1}^{N} \delta_i^2 \sum_{i=1}^{N} V_i,$$

where $N$, $B$, $\delta_i$, $V_i$ are the number of atoms of a mesh, the Bulk modulus of Au, the atomic dilation and Voronoi volume, respectively. The atomic dilation $\delta$ was obtained by

$$\delta = \frac{s_{xx} + s_{yy} + s_{zz}}{E},$$

whence ($s_{xx}$, $s_{yy}$, $s_{zz}$) are the normal components of the atomic virial stress tensor and $E$ is Young's modulus of Au (79 GPa). The atomic virial stress tensor was obtained from the '*stress/atom*' normalized by its Voronoi volume in the LAMMPS and averaged over 50 time-steps.

**5. Relaxation of the assembled NC structure using molecular dynamics**

The assembled NC structure was optimized as follows using LAMMPS[9]. To remove unphysical atomic arrangements such as abrupt interfaces at the mesh boundaries, the Au atoms within 0.36 nm from the mesh boundaries were first excited to 300 K by adding velocities, while the rest of the atoms remained at rest. Next, the entire Au NP was relaxed for 400 fs in a microcanonical ensemble using the Embedded Atom Method (EAM) potential of Grochola *et al*[10]. This EAM potential was generated by fitting to high-temperature lattice constants and liquid densities, and thus is relevant to relax the initial model, which involves strong repulsion at mesh boundaries. Finally, the system was further relaxed until the temperature converged using the EAM potential of Foiles *et al*[11]. Atomic structure model for the ideal TOH Au NC was optimized by heating the entire NC to 300 K and then relaxing for 2 ps using the EAM potential of Foiles *et al.* alone, as this ideal model involved no patching of small volumes.

## Supplementary References

1 Ayyer, K., Lan, T.-Y., Elser, V. & Loh, N. D. Dragonfly: an implementation of the expand-maximize-compress algorithm for single-particle imaging. *J. App. Cryst.* **49**, 1320-1335 (2016).

2 Ekeberg, T. *et al.* Three-Dimensional Reconstruction of the Giant Mimivirus Particle with an X-Ray Free-Electron Laser. *Phys. Rev. Lett* **114**, 098102 (2015).

3 Loh, N.-T. D., Eisebitt, S., Flewett, S. & Elser, V. Recovering magnetization distributions from their noisy diffraction data. *Phys. Rev. E* **82**, 061128 (2010).

4 Kruskal, W. H. & Wallis, W. A. Use of Ranks in One-Criterion Variance Analysis. *Journal of the American Statistical Association* **47**, 583-621, doi:10.2307/2280779 (1952).

5 Pham, M., Yin, P., Rana, A., Osher, S. & Miao, J. Generalized proximal smoothing (GPS) for phase retrieval. *Opt. Express* **27**, 2792-2808 (2019).

6 Rodriguez, J. A., Xu, R., Chen, C.-C., Zou, Y. & Miao, J. Oversampling smoothness: an effective algorithm for phase retrieval of noisy diffraction intensities. *J. App. Cryst.* **46**, 312-318 (2013).

7 Marchesini, S. *et al.* X-ray image reconstruction from a diffraction pattern alone. *Phys. Rev. B* **68**, 140101 (2003).

8 Ayyer, K. *et al.* Low-signal limit of X-ray single particle diffractive imaging. *Opt. Express* **27**, 37816-37816, doi:10.1364/OE.27.037816 (2019).

9 Plimpton, S. Fast Parallel Algorithms for Short-Range Molecular Dynamics. *J. Comp. Phys.* **117**, 1-19 (1995).

10 Grochola, G., Snook, I. K. & Russo, S. P. Computational modeling of nanorod growth. *J. Chem. Phys.* **127**, 194707 (2007).

11 Foiles, S. M., Baskes, M. I. & Daw, M. S. Embedded-atom-method functions for the fcc metals Cu, Ag, Au, Ni, Pd, Pt, and their alloys. *Phys. Rev. B* **33**, 7983-7991 (1986).

12 Gamaly, E. G. *Femtosecond laser-matter interaction: theory, experiments and applications*. 92 (CRC Press, 2011).